\documentclass[a4paper,preprint,superscriptaddress,preprintnumbers,nofootinbib]{revtex4}

\usepackage[dvipdfmx]{graphicx}
\usepackage{amsmath}
\usepackage{amssymb}
\usepackage{float}
\usepackage{wrapfig}
\usepackage{bm}
\usepackage{color}

\newcommand{\Slash}[1]{{\ooalign{\hfil/\hfil\crcr$#1$}}}

\newcommand{\e}{{\rm e}}

\newcommand{\D}{\mathcal{D}}

\newcommand{\al}[1]{\begin{align}#1\end{align}}
\newcommand{\bp}{\begin{pmatrix}}
\newcommand{\ep}{\end{pmatrix}}
\newcommand{\nn}{\nonumber\\}

\newcommand{\p}{\partial}

\newcommand{\df}{\text{d}}

\newcommand{\bs}[1]{\boldsymbol}

\newcommand{\pmat}[1]{\begin{pmatrix}#1\end{pmatrix}}

\newcommand{\fn}[1]{\!\left(#1\right)}

\newcommand{\bra}{\langle}
\newcommand{\ket}{\rangle}

\begin{document}
\title{Phase structure of NJL model with weak renormalization group}

\author{Ken-Ichi \surname{Aoki}}
\email{aoki@hep.s.kanazawa-u.ac.jp}
\affiliation{Institute for Theoretical Physics, Kanazawa University, Kanazawa 920-1192, Japan}

\author{Shin-Ichiro \surname{Kumamoto}}
\email{kumamoto@rieb.kobe-u.ac.jp}
\affiliation{Research Institute for Economics and Business Administration, Kobe University, Kobe 657-8501, Japan}

\author{Masatoshi \surname{Yamada}}
\email{m.yamada@thphys.uni-heidelberg.de}
\affiliation{Institut f\"ur Theoretische Physik, Universit\"at Heidelberg, Philosophenweg 16, 69120 Heidelberg, Germany}

\preprint{KANAZAWA-17-04}

\begin{abstract}
We analyze the chiral phase structure of the Nambu--Jona-Lasinio model at finite temperature and density by using the functional renormalization group (FRG).
The renormalization group (RG) equation for the fermionic effective potential $V(\sigma;t)$ is given as a partial differential equation, where $\sigma:=\bar \psi\psi$ and $t$ is a dimensionless RG scale.
When the dynamical chiral symmetry breaking (D$\chi$SB) occurs at a certain scale $t_\text{c}$, $V(\sigma;t)$ has singularities originated from the phase transitions, and then one cannot follow RG flows after $t_\text{c}$. 
In this study, we introduce the weak solution method to the RG equation in order to follow the RG flows after the D$\chi$SB and to evaluate the dynamical mass and the chiral condensate in low energy scales.
It is shown that the weak solution of the RG equation correctly captures vacuum structures and critical phenomena within the pure fermionic system.
We show the chiral phase diagram on temperature, chemical potential and the four-Fermi coupling constant.
\end{abstract}
\maketitle

\section{Introduction}
Solving Quantum Chromodynamics (QCD) is a challenging problem in elementary particle physics.
Due to the strong dynamics of QCD in macroscopic scales, the phenomena such as the quark confinement and the dynamical chiral symmetry breaking (D$\chi$SB) are predicted.
The exploration of the phase structure and the equation of state in hot and dense QCD are quite important for understanding the evolution of quark matters in the universe.

Numerous studies have been performed and have shown that QCD at finite temperature and density has various phase structures, e.g., the quark-gluon plasma and the color superconductivity.
In particular, the study by lattice Monte Carlo simulation has substantially contributed to the understanding of nature of QCD.
At present, lattice QCD, however, suffers from the sign problem.
Since the Boltzmann weight at finite density has imaginary parts, the generated configurations oscillate and thus the statistical errors become large.

In contrast, the analytical methods, e.g., the mean-field approximation (MFA) and the Schwinger--Dyson equation (SDE) with the ladder approximation, have been applied to QCD and have succeeded to understand its dynamics and phase structures both qualitatively and quantitatively.
Furthermore, effective theory approaches have also helped us to understand QCD.
The Nambu--Jona-Lasinio (NJL) model~\cite{Nambu:1961tp,Nambu:1961fr} and its extended models~\cite{Fukushima:2003fw,Fukushima:2008wg}, especially, have also played crucial roles as effective models describing the D$\chi$SB and the confinement in QCD; see also~\cite{Hatsuda:1994pi,Du:2015psa,Rozynek:2015zca}.
The Lagrangian of the NJL model is given by the following four-Fermi interaction structure:
\al{
{\mathcal L}_{\rm NJL}= \bar\psi i\Slash \p \psi -\frac{G_0}{2}\left[ (\bar \psi \psi)^2 + (\bar \psi i\gamma^5\psi)^2 \right].
\label{NJLmodelLag}
}
This Lagrangian is invariant under the chiral $\text{U}\fn{1}$ transformation: $\psi \to \e^{i\gamma^5\theta}\psi$ and $\bar \psi \to \bar \psi \e^{i\gamma^5 \theta}$.
Such a four-Fermi interaction would be generated via the interaction between gluon and quark fields within QCD; see e.g., \cite{Mitter:2014wpa}.
For a larger coupling constant $G_0$ than a certain critical coupling constant $G_c$, the non-trivial vacuum, i.e., the chiral condensate $\bra \bar \psi \psi \ket$, is generated.

Let us here consider the analysis of the D$\chi$SB in the NJL model at finite temperature and density with the functional renormalization group (FRG) method~\cite{Aoki:1999dw,Aoki:1999dv,Kodama:1999if,Aoki:2000dh,Braun:2011pp,Braun:2011fw,Braun:2012zq,Jakovac:2013jua,Aoki:2014ova,Aoki:2015hsa,Braun:2017srn};
see also \cite{Aoki:2000wm,Bagnuls:2000ae,Berges:2000ew,Polonyi:2001se,Pawlowski:2005xe,Gies:2006wv,Rosten:2010vm} as the reviews of the FRG.
The advantage of the FRG is that one can systematically improve approximations for the FRG analysis.
However, as will be explained in the next section, the renormalization group (RG) equation of the NJL model with the strong interaction has singularities originated from the divergence of four-Fermi coupling constant.
Therefore, it is difficult to follow the solution of the RG equation after arising the singularities.
The method, which have been employed in order to overcome the singularities, is the auxiliary field method based on the Hubbard--Stratonovich transformation~\cite{Aoki:1999dv}.\footnote{
Even if the auxiliary field is introduced at the initial scale, the four-Fermi interactions are generated in the middle of scale by the box diagrams made of the Yukawa interaction between the quark and the auxiliary field.
To avoid it, the dynamical hadronization method (or rebosonization) has been developed and used to investigate the vacuum structure of QCD~\cite{Gies:2001nw,Gies:2002hq,Jaeckel:2002rm,Pawlowski:2005xe,Gies:2006wv,Braun:2008pi,Floerchinger:2009uf,Mitter:2014wpa,Braun:2014ata,Rennecke:2015eba,Fu:2015naa}.
}
In this method, the four-Fermi vertex is replaced by exchanges of the bosonic auxiliary field $\phi$ via the Yukawa interactions $y\phi \bar \psi \psi$, namely, $G_0\sim y^2/m^2$.\footnote{
Note that it may seem that the number of free parameters increases from one to two (from $G_0$ to $y$ and $m^2$).
However, one of them should be redundant.
It has been shown in \cite{Aoki:1999dv,Mitter:2014wpa,Braun:2014ata} that the low energy physics actually does not depend on the initial value of the Yukawa coupling constant.
}
Using this method one can not only avoid the singularities but also include the mesonic quantum effects.
However, analyses of the system becomes complicated due to the additional bosonic degrees of freedom and ambiguities of approximations within the FRG method.
Although the method of bare mass is also used to system~\cite{Aoki:2009zza}, the operator expansion for the effective potential~\eqref{expansion of potential} does not converge in the small bare mass limit, and then it is difficult to obtain physical values in the chiral limit~\cite{Aoki:2012mj}.
Moreover, the most serious problem in the method of bare mass is that one cannot address the chiral phase transition of first-order.

Recently, to overcome the singularities, a novel method called the weak solution method was suggested~\cite{Aoki:2014ola}.
The solution of the RG equation with the singularities is mathematically defined as the ``weak'' solution.
Hence, we call the RG equation applied the weak solution method ``weak renormalization group equation''.
It has been shown in \cite{Aoki:2014ola} that the weak solution method actually can address the chiral phase transitions of both second- and first-order and evaluate the physical values such as the dynamical mass and the chiral condensate.

In this paper, we analyze the D$\chi$SB in the NJL model at finite temperature and density using the FRG with the weak solution method.
The chiral phase diagram of the NJL model will be investigated.
It should be emphasized that we will obtain the chiral phase diagram of the NJL model using the FRG without the auxiliary field method.
Previously, only the auxiliary field method within the FRG has been able to capture the first-order phase transition.
The main purpose of this paper is to show that one can obtain the chiral phase diagram with both second- and first-order phase transitions by the weak solution method without the help of the auxiliary field method. 

This paper is organized as follows:
In the section \ref{njl model and weak nprg}, we introduce the model and its effective model. The FRG method is briefly explained in order to apply it to the model.
In the section \ref{weak solution}, we introduce the weak solution method and give the weak RG equation.
The numerical results are exhibited in Sec.~\ref{numerical analysis}.
We summarize and discuss our approach to the system and the results in Sec.~\ref{summary and discussion}.
In the appendix~\ref{derivation of beta function}, we show how to derive the RG equation.
In the appendix~\ref{concavity and convexity}, the convexity and concavity of the beta function is discussed.

\section{Nambu--Jona-Lasinio model and weak renormalization group}\label{njl model and weak nprg}
Our purpose is to investigate the D$\chi$SB described by the NJL model in $d=4$ and the chiral phase transition at finite temperature and density.
In this section, we give the action of the NJL model and the RG equation for the effective potential.
We see that the effective potential in the NJL model becomes singular within the RG evolution.

\subsection{NJL model}
The Lagrangian of the NJL model with one flavor is given in Eq.~\eqref{NJLmodelLag}.
Since in this paper we would like to investigate the fundamental behavior of the D$\chi$SB by using the weak RG, we consider the following simplified NJL model with one flavor in four dimension Euclidean space:
\al{\label{njl action}
S_{\rm NJL}= \int \df^4x\left[\bar\psi \Slash \p \psi -\frac{G_0}{2}(\bar \psi \psi)^2\right],
}
where we assume that the quark fields have the degree of freedom of color, $N_\text{c}$.
Since the pseudo-scalar operator $(\bar \psi i\gamma^5 \psi)^2$ is not introduced, this action has no continuous chiral symmetry.
Instead, it is invariant under the discrete type transformation,
\al{
\psi &\to \gamma^5 \psi,&
\bar \psi &\to - \bar \psi \gamma^5,&
}
which forbids the operators with negative power of $\bar \psi \psi$. 
Hence, the mass term of fermion $m\bar \psi \psi$ does not appear in the action.
This model describes that the chiral symmetry breaks down due to the strong four-Fermi interaction and the non-trivial vacuum, i.e., the non-zero chiral condensate $\bra \bar \psi \psi \ket \neq 0$, is realized.
In this vacuum, the fermion obtains the dynamical mass which is proportional to the chiral condensate.
It should be noted that we do not really work for the discrete chiral symmetric model with only scalar four-fermi interactions.
Starting with the continuous chiral symmetric model with scalar and pseudo-scalar four-fermi interactions together, we define the renormalized fermionic interactions expressed in terms of 
scalar bilinear $\bar\psi \psi$ and pseudo-scalar bilinear $\bar\psi i\gamma_5\psi$ operators. 
The renormalized interactions have the same continuous chiral symmetry, since the RG beta functions are set to respect the symmetry.

After integrating the RG equation to the infrared end, we investigate the free energy (vacuum amplitude) as a function of scalar and pseudo-scalar source fields. The free energy is readily obtained making use of the translational invariance feature of our beta functions. 
By the Legendre transformation in a usual way, we define the effective potential as a function of 
$\sigma=\langle\bar\psi\psi\rangle$ and $\pi=\langle\bar\psi i\gamma_5\psi\rangle$, which has the same chiral symmetric form, i.e., it is a function of $\sigma^2 + \pi^2$.

Therefore only a section of the effective potential where $\pi=0$ is sufficient to get the radial profile of it.
It is shown straightforwardly that this section does not depend on the coupling constant of elementary pseudo-scalar interactions in the bare Lagrangian.
Then we set the vanishing pseudo-scalar coupling constant.
This is our strategy of model set up, which gives the shortest way of evaluating the effective potential profile.\footnote{\label{comment on flavor case}
We comment on the finite number flavor case.
In a general flavor case, the chiral symmetry is given as $\text{SU}\fn{N_\text{f}}_V\times \text{SU}\fn{N_\text{f}}_A\times \text{U}\fn{1}_V\times \text{U}\fn{1}_A$.
$ \text{U}\fn{1}_A$ symmetry is broken by the chiral anomaly.
This effective is described by the 't Hooft--Kobayashi--Maskawa term~\cite{Kobayashi:1970ji,tHooft:1976snw,tHooft:1986ooh} within the effective model approach.
Since this term with $N_\text{f}=2$ is given as the four-Fermi interaction, we would expect no crucial difference from the present setup.
In contrast, in the $N_\text{f}=3$ (or $2+1$) case, 't Hooft--Kobayashi--Maskawa term is given by the six-Fermi interaction.
Although the anomaly effect on the chiral phase transition is investigated in several literatures, e.g. \cite{Lenaghan:2000ey,Roder:2003uz,Mitter:2013fxa}, it might depend on chiral effective models.
}
We should, however, take account of the pseudo-scalar operator to calculate, for example, critical exponents, which are sensitive to the difference of the discrete and continuous symmetries.

\subsection{FRG equation}
We introduce the FRG to investigate the D$\chi$SB in the NJL model.
Now, we briefly explain its idea and derivation.
Let us consider the bare action $S[\phi;\Lambda_0]$ and the path integral,
\al{\label{path integral}
Z=\int \D \phi\, \e^{-S[\phi;\Lambda_0]}.
}
This integration can be interpreted as the summation of the quantum fluctuations with the momentum $0<|p|<\Lambda_0$.
In the FRG, the path integral is evaluated by integrating out the quantum fluctuation with the higher momentum $\Lambda<|p|<\Lambda_0$, namely,
\al{
Z
=\int \D \phi_< \D \phi_>\, \e^{-S[\phi_<+\phi_>;\Lambda_0]}
=\int \D \phi_<\, \e^{-S_{\rm eff}[\phi_<;\Lambda]},
}
where we divided the field $\phi\fn{p}$ into the higher momentum mode $\phi_>\fn{p}$ with $\Lambda<|p| <\Lambda_0$ and the lower momentum mode $\phi_<\fn{p}$ with $0<|p| <\Lambda$.
The action $S_{\rm eff}[\phi_<;\Lambda]$ is called the Wilsonian effective action  and is defined by the lower momentum mode.
The FRG describes the change from the bare action $S[\phi;\Lambda_0]$ to the effective action $S_{\rm eff}[\phi_<;0]$ through integrating the shell momentum $\Lambda-\delta\Lambda<|p|<\Lambda$, which is the so-called coarse graining.
This step is formulated as a functional differential equation, and solving this equation is equivalent to evaluating the path integral~\eqref{path integral}.
In this paper, we employ the Wegner-Houghton (WH) equation~\cite{Wegner:1972ih},
\al{\label{wegner houghton eqaution}
\frac{\df S_{\rm eff}}{\df t}=-\frac{1}{2} \int_{\rm shell} \frac{\df ^dp}{(2\pi)^d}
\left\{
\frac{\delta S_{\rm eff}}{\delta \phi_p}
\left( \frac{\delta^2 S_{\rm eff}}{\delta \phi_p\delta \phi_{-p}} \right)^{-1}
\frac{\delta S_{\rm eff}}{\delta \phi_{-p}}
-{\rm str}\log\fn{ \frac{\overrightarrow \delta}{\delta \phi_p} S_{\rm eff}\frac{\overleftarrow \delta}{\delta \phi_{-p} }}
\right\},
}
where $\phi_p:=\phi\fn{p}$ and we introduced the dimensionless scale parameter,
\al{
t:=\log\fn{\Lambda_0/\Lambda}.
}
It is rewritten as $\Lambda =\Lambda_0\e^{-t}$, and then the increasing $t$ corresponds to the decreasing the cutoff $\Lambda$.
Since  the two point function $\frac{\overrightarrow \delta}{\delta \phi_p} S_{\rm eff}\frac{\overleftarrow \delta}{\delta \phi_{-p} }$ is generally given as the supermatrix in field space, ``str'' denotes taking the trace for the supermatrix; see~\cite{Clark:1992jr,Oda:2015sma} for the treatments of the supermatrix and its manipulations.
For details of derivation of the WH equation, see e.g.~\cite{Clark:1992jr}.\footnote{
The WH equation is also obtained from the sharp cutoff limit of the Plochinski equation~\cite{Polchinski:1983gv,Sumi:2000xp}
}
The Wilsonian effective action is generally spanned by an infinite number of effective operators generated by the interactions in the bare action. 
Note that symmetries which the bare action has are not broken through the shell momentum integral, hence, the generated effective operators respect their symmetries.

Although the WH equation~\eqref{wegner houghton eqaution} itself is exact, we cannot solve it without some approximations.
That is, we have to restrict the theory space of the Wilsonian effective action to its subspace.
We give the following action as the approximated Wilsonian effective action for the NJL model~\eqref{njl action}:
\al{\label{effective njl action}
S_{\rm eff}[ \psi, \bar\psi;\Lambda]
= \int \df^4 x\left[\bar\psi \Slash \p \psi
 -V\fn{\bar\psi \psi;\Lambda}\right].
}
If the effective potential is expanded into the polynomials of $\bar \psi \psi$, we have
\al{\label{expansion of potential}
V\fn{\bar\psi \psi;\Lambda}
=\frac{G_\Lambda}{2}(\bar\psi \psi)^2 +\frac{G_{4,\Lambda}}{4}(\bar\psi \psi)^4 +\cdots.
}
We call this potential the ``fermionic" effective potential.
As mentioned above, the odd powers of $\bar \psi \psi$ is not generated due to the (discrete) chiral symmetry. 
The potential at the initial scale $\Lambda_0$ is given by 
\al{
V\fn{\bar \psi \psi; \Lambda_0}
=\frac{G_0}{2}\left( \bar \psi \psi \right)^2,
}
that is, $G_{\Lambda_0}\equiv G_0$ and other effective coupling constants vanish.
Hereafter, we will use the notations $\sigma:=\bar\psi \psi$.
Note that the variable $\sigma$ is not an auxiliary field, accordingly its derivatives with respect to $\psi$ and $\bar \psi$ are given as
\al{
\sigma\, \frac{\overleftarrow\p}{\p \psi}&=\bar \psi,&
\frac{\overrightarrow \p }{\p \bar \psi}\,\sigma&=\psi.&
}

We will discuss the chiral phase transition by the effects of temperature and density in the next section.
For this purpose, the density operator $\int \df ^4x \, \mu \bar \psi \gamma_0\psi$ should be introduced in the bare action~\eqref{njl action} and the Wilsonian effective action~\eqref{effective njl action}.\footnote{
We assume that the chemical potential $\mu$ is not corrected by quantum fluctuations.
}
Moreover, the momentum in the time direction $p_0$ is replaced with the Matsubara frequency $\omega_n$ and its integration becomes the Matsubara summation.
It is complicated to evaluate the shell momentum integral for four dimension momentum.
To avoid it, we insert the cutoff $\Lambda$ into the momentum in the spatial direction.
Then the shell momentum integral becomes
\al{\label{shell mode 3d}
\int_{\rm shell}\frac{\df ^4p}{(2\pi)^4} 
=T\displaystyle \sum_{n=-\infty}^\infty \int \frac{\df ^{3}p}{(2\pi)^{3}}\,\Lambda\,\delta\fn{|\vec p|-\Lambda}.
}

The RG equation for the effective potential is given by
\al{\label{dimensionful RG equation}
\p_t V\fn{\sigma;t}= \frac{\Lambda^3}{\pi^2}
\bigg[
E +T\log \fn{1+\e^{-\beta (E-\mu)}} +T \log \fn{1+\e^{-\beta(E+\mu)}}
\bigg],
}
where $\beta=1/T$, $E:= \sqrt{\Lambda ^2 +M^2}$ and $M:=\p_\sigma V$.
We show the explicit derivation of the RG equation in the appendix~\ref{derivation of beta function}.
Note that the large-$N$ approximation ($N_\text{c} \to \infty$) and the uniform $\bar{\psi}\psi$ approximation were employed in order to obtain Eq.~\eqref{dimensionful RG equation}.
The function $M$ corresponds to the dynamical mass of quark.
We here call it the mass function.

Here, we revisit the NJL model at vanishing temperature and density.
The RG equation~\eqref{dimensionful RG equation} at zero temperature and zero density becomes
\al{\label{zero temp rg equation}
\p_tV\fn{\sigma;t}= \frac{\Lambda^3}{\pi^2}\sqrt{\Lambda^2 +\left( \p_\sigma V\right)^2}.
}
Using the expansion of the effective potential given in Eq.~\eqref{expansion of potential}, we obtain the RG equations of each coupling constant. 
In particular, the RG equation of the four-Fermi coupling constant is given by 
\al{
\p_t G =  \frac{G^2\Lambda^2}{\pi^2}.
}
If we use the dimensionless rescaled coupling constant $\tilde G:= G\Lambda^2/2\pi^2$,
it becomes
\al{\label{dimensionless rescaled four-fermi coupling constant}
\p_t \tilde G = -2\tilde G +2\tilde G^2.
}
Obviously, this equation has the fixed point $\tilde G_{\rm c}=1$ and yields the blow-up solution,
\al{
\tilde G\fn{t}=\frac{\tilde G_{\rm c} \tilde G_0}{\tilde G_0 -(\tilde G_0-\tilde G_{\rm c})\e^{2t}},
}
for the initial value $\tilde G_0> \tilde G_{\rm c}$.
We see the critical scale at which the denominator becomes zero,
\al{
t_{\rm c}= \frac{1}{2}\log \fn{\frac{\tilde G_0}{\tilde G_0-\tilde G_{\rm c}}}.
}
The divergence of $\tilde G$ itself is physically relevant since it corresponds to the chiral fluctuation $\tilde G \sim\bra (\bar\psi \psi)^2\ket$.
Thus, the divergence is the signal of the second order phase transition.
However, the divergence disturbs solving the RG equation up to the infrared (IR) scale $\Lambda \to 0$.
Since the four-Fermi coupling constant is lowest order of the effective potential~\eqref{expansion of potential}, the effective potential becomes non-differentiable at its origin $V\fn{\sigma=0}$ due to the singularity of the four-Fermi coupling constant. 
In other words, the derivative $\p_\sigma V$ cannot be defined after the critical scale $t_{\rm c}$.

We see in the next section that the weak solution method~\cite{Aoki:2014ola} allows us to analyze the D$\chi$SB without introducing the auxiliary field and to investigate the chiral first-order phase transition.

\section{Weak solution method}\label{weak solution}
In this section, the weak solution of the RG equation \eqref{dimensionful RG equation} is defined.
To construct it numerically, we introduce the method of characteristics which are given as the system of ordinary differential equations made from the partial differential equation.
Moreover, to obtain a single-valued solution from a multi-valued solution, we use the Rankine--Hugoniot (RH) condition which are introduced by the definition of the weak solution.
It is demonstrated how the solutions of characteristics with the RH condition choose a unique vacuum.
Finally, we comment on the RG equation for the four-Fermi coupling constant from the viewpoint of the weak solution method. 
\subsection{Definition of weak solution}
As discussed in the previous section, the RG equation for the effective potential becomes singular when the D$\chi$SB occurs.
In \cite{Aoki:2014ola}, it was shown that the weak solution method is valid for overcoming the singularity of this system.
Here, we briefly review basic notions of the weak solution method and show the weak RG equation for Eq.~\eqref{dimensionful RG equation}.

The RG equation \eqref{dimensionful RG equation} is given as the partial differential equation,
\al{\label{beta function v}
\p_t V\fn{\sigma;t}=-F\fn{M\fn{\sigma;t},\sigma;t},
}
where $F\fn{M,\sigma;t}$ is the beta function with a negative sign.
Differentiating the both sides of RG equation with respect to $\sigma$, we obtain
\al{\label{mass rg}
\p_tM\fn{\sigma;t}+\p_\sigma F\fn{M\fn{\sigma;t},\sigma;t}=0.
}
Let us now call this equation the ``strong'' RG equation in order to compare with the ``weak'' RG equation introduced below. 

Here, we introduce a test function $\varphi\fn{\sigma;t}$ which is smooth and converges to zero for both infinite $t$  and $\sigma$: $\varphi\fn{\pm \infty;t}=\varphi\fn{\sigma;\infty}=0$.
Integrating Eq.~\eqref{mass rg} multiplied by this function, we have
\al{
\int ^\infty_0\df t \int ^{\infty}_{-\infty}\df \sigma 
\left( \p_tM+\p_\sigma F \right) \varphi\fn{\sigma;t}=0.
}
By utilizing the integration by parts, it becomes
\al{\label{weak equation}
\int ^\infty_0\df t \int ^{\infty}_{-\infty}\df \sigma
\left( M\frac{\p \varphi}{\p t} +F \frac{\p \varphi}{\p \sigma}\right)
+\int^{\infty}_{-\infty} \df \sigma\, (M\, \varphi)|_{t=0}=0.
}
We call the RG equation~\eqref{weak equation} ``weak RG equation", and its solution is called ``weak solution''.
The derivatives with respect to $t$ and $\sigma$ move from the mass function and the beta function to the test function.
Hence, $M\fn{\sigma;t}$ given as a solution of the weak RG equation~\eqref{weak equation} can have an arbitrary number of singularities at any point on the $\sigma$--$t$ plane.
\subsection{Method of characteristics}
The RG equation of the mass function~\eqref{mass rg} is given as a partial differential equation.
Here, we introduce the method of characteristics which reduces from the partial differential equation to the system of ordinary differential equations.
The method of characteristics allows us to numerically solve the RG equation using e.g., the Runge-Kutta method and the finite-difference methods.

Let us consider a curved surface $M\fn{\sigma;t}$ in the $\sigma$--$t$--$z$ space and rewrite the RG equation~\eqref{mass rg} as 
\al{
\pmat{
\p_M F & 1 & 0
}
\pmat{
\p_\sigma M\\
\p_t M\\
-1
}
=0,
}
where we used $\p_\sigma F=(\p_M F)(\p_\sigma M)$.
The infinitesimal displacement vector $(\df \sigma, \df t, \df M)$ is a tangent vector for a point $(\sigma,t,M)$ on the curved surface $M\fn{\sigma;t}$, and satisfies
\al{
\pmat{
\df \sigma & \df t & \df M
}
\pmat{
\p_\sigma M\\
\p_t M\\
-1
}
=0,
}
since the total derivative of $M$ is given by $\df M = \p_\sigma M\, \df \sigma + \p_t M\, \df t$.
Thus, the vector $(\p_\sigma M~\p_t M\,-1)^{\rm T}$ is orthogonal to $(\df \sigma~\df t~ \df M)$, and we find that the vector $(\p_M F ~ 1~  0)$ is also the tangent vector for a point on the curved surface $M\fn{\sigma;t}$.
Taking $(\df \sigma~ \df t~ \df M)$ as the vector being proportional to $(\p_M F ~1~ 0)$, we have
\al{
&\frac{\df \sigma \fn{s}}{\df s}=\frac{\p F}{\p M},\label{carac1}\\
&\frac{\df t\fn{s}}{\df s}=1,\label{carac2}\\
&\frac{\df M\fn{\sigma ,s}}{\df s}=0\label{carac3},
}
where we introduced an infinitesimal parameter $\df s$ as a proportionality constant.
The solutions $\left( \sigma\fn{s},t\fn{s},M\fn{s} \right)$ of the system of ordinary differential equations with each initial conditions are called characteristics.
Since the solution of Eq.~\eqref{carac2} is simply $t=s$, we can replace from $s$ to $t$ in Eqs.~\eqref{carac1} and \eqref{carac3}.

The fermionic effective potential $V$ can be evaluated within the method of characteristics.
Considering $V$ on the characteristics, we obtain
\al{
\frac{\df V\fn{s}}{\df s}
&=\frac{\p V\fn{\sigma;s}}{\p \sigma}\frac{\df \sigma\fn{s}}{\p s}
+\frac{\p V\fn{\sigma;s}}{\p s}\nn
&=M\frac{\p F}{\p M}-F,
}
where we used Eqs.~\eqref{mass rg}, \eqref{carac1} and $M=\p_\sigma V$. 

To summarize, we have the following coupled equations for the present system:
\al{
&\frac{\df \sigma \fn{t}}{\df t}=\frac{\p F}{\p M},&
&\frac{\df M\fn{\sigma ,t}}{\df t}=0,&
&\frac{\df V\fn{t}}{\df t}=M\frac{\p F}{\p M}-F.\label{caracta3}&
}
Actually, these equations are independent of each other, therefore, their solutions are given by
\al{
&M\fn{\sigma\fn{t};t}=M\fn{\sigma_0;0}=:M_0
\label{cha1}
,\\
&\sigma\fn{t}=\sigma_0+\int ^t_0\df \tau \, \p_M F\fn{M_0;\tau}
\label{cha2}
,\\
&V\fn{\sigma;t}= V\fn{\sigma_0;0}
+\int ^t_0 \df \tau\, \left. \left[ M\frac{\p F}{\p M} -F \right]\right|_{M=M_0},
\label{cha3}
} 
where $\sigma_0:=\sigma\fn{0}$.
Note here that $F\fn{M,\sigma;t}$ given in the right-hand side of Eq.~\eqref{dimensionful RG equation} (with negative sign) does not explicitly depend on $\sigma$; $F\fn{M,\sigma;t}\equiv F\fn{M;t}$.
We also note that this simplification happens in the NJL model with a certain approximation.
Although for more general cases such as QCD, we have to solve the system of ordinary differential equations for a given system, solving it is easier than the case of the partial differential equation. 

\begin{figure}
\begin{center}
\includegraphics[width=100mm]{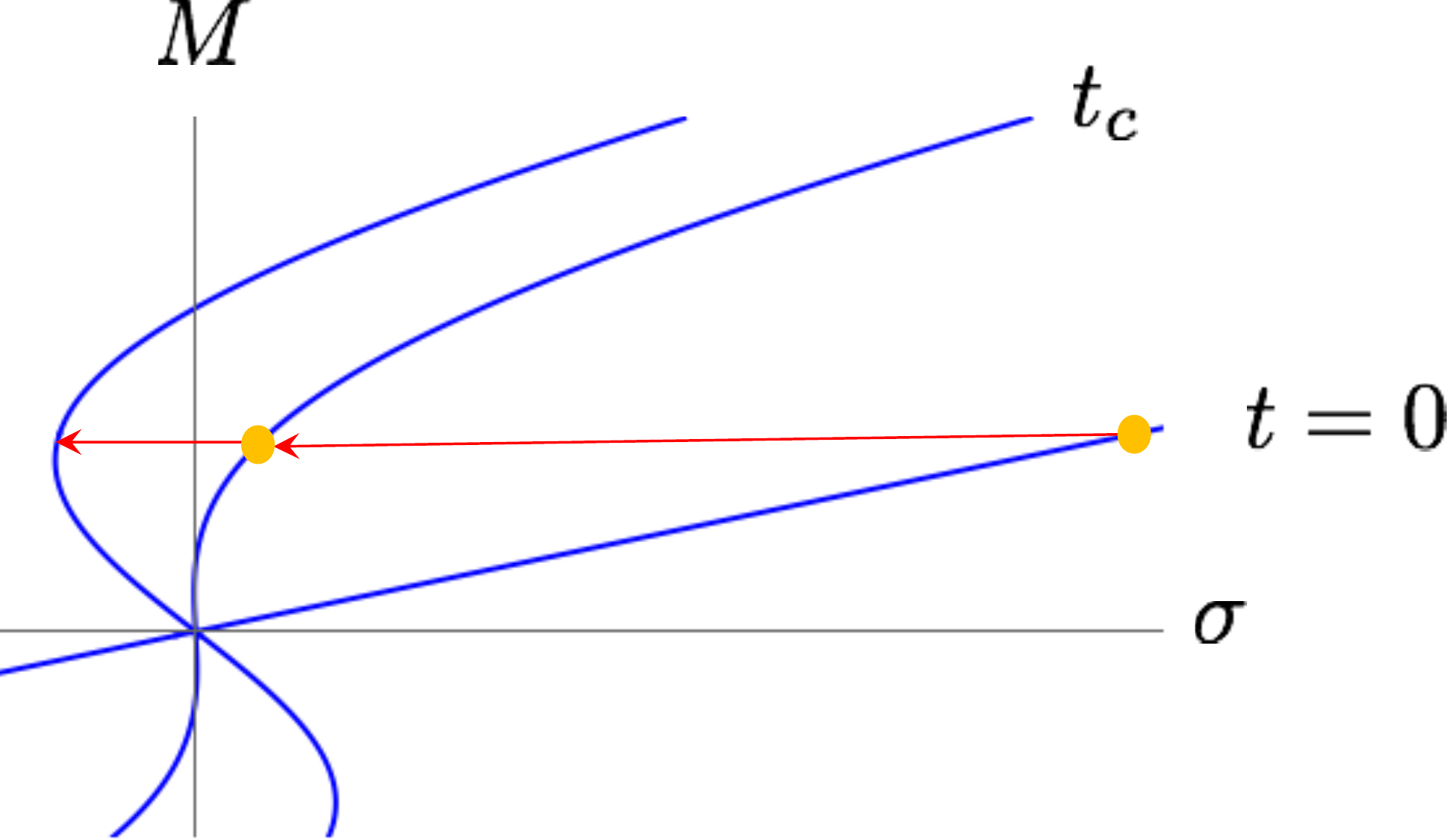}
\end{center}
\caption{Schematic figure for the solution of Eqs.~\eqref{cha1}--\eqref{cha3}.
Note that the initial mass function is given by $M\fn{\sigma;t=0}=G_0 \sigma$.
}
\label{weak_schematic_motion}
\end{figure}

A schematic figure for the evolution of the mass function described by Eqs.~\eqref{cha1}--\eqref{cha3} is shown in Fig.~\ref{weak_schematic_motion}.
Note that for a fixed $M_0>0$, $\sigma$ described by Eq.~\eqref{cha2} moves towards the left-hand side since $\p^2F/\p M^2<0$ for Eq.~\eqref{dimensionful RG equation} is satisfied; see the appendix~\ref{concavity and convexity}.
The system of ordinary differential equations \eqref{cha1}--\eqref{cha3} does not yield any singularity.
Instead, as a schematic figure shown in Fig.~\ref{weak_schematic}, after the critical scale $t_c$, the mass function becomes a multi-valued function around a value $\sigma^*$, at which the fermionic effective potential becomes the swallowtail form (the right-hand side of Fig.~\ref{weak_schematic}).
However, the solution of the fermionic potential and the mass function after the D$\chi$SB have to be a single-valued function of $\sigma$ since there should exist only one physical vacuum.
To uniquely determine the physical vacuum, we give the Rankine--Hugoniot condition in the next subsection.

\subsection{Rankine--Hugoniot condition}
\begin{figure}
\begin{center}
\includegraphics[width=140mm]{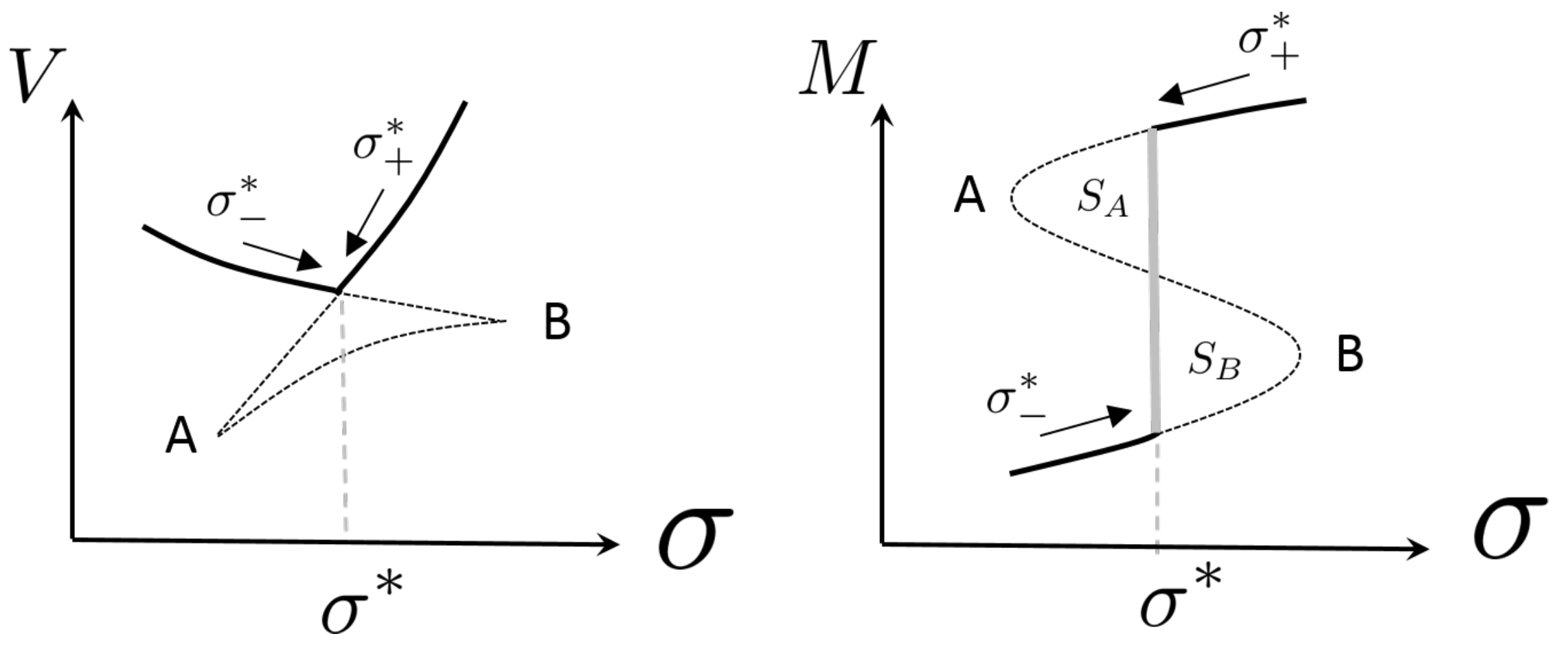}
\end{center}
\caption{Schematic figures for the effective potential $V$ (left) and the mass function $M$ (right) after the D$\chi$SB.}
\label{weak_schematic}
\end{figure}
To uniquely determine the weak solution from Eq.~\eqref{weak equation} (or equivalently Eq.~\eqref{caracta3}), we give a condition called the Rankine--Hugoniot (RH) condition. 
This condition is derived from the definition of the weak solution and reads
\al{
\frac{\df \sigma^*\fn{t}}{\df t}
=\frac{F\fn{M\fn{\sigma_+^*\fn{t}};t}-F\fn{M\fn{\sigma_-^*\fn{t};t}}}{M\fn{\sigma_+^*\fn{t};t}-M\fn{\sigma_-^*\fn{t};t}},
}
where $\sigma^*\fn{t}$ is a discontinuity point, and $\sigma_+^*$ and $\sigma_-^*$ denote closing to the point $\sigma^*$ from the right-hand side ($+$) and the left-hand side ($-$), respectively (see Fig.~\ref{weak_schematic}).
This condition describes the evolution of $\sigma^*\fn{t}$ and implies geometrically the differential-area invariance, namely,
\al{
S_A\fn{t}-S_B\fn{t}={\rm constant}.
\label{RHc}
} 
In the present case, since at the critical scale $t=t_\text{c}$, $S_A\fn{t_\text{c}}=S_B\fn{t_\text{c}}=0$ is satisfied, the constant value in Eq.~\eqref{RHc} vanishes, and then $S_A\fn{t}=S_B\fn{t}$.
In Fig.~\ref{weak_schematic}, the RH condition means that the vertical line (gray line) is defined such that the equal area rule $S_A\fn{t}=S_B\fn{t}$ is satisfied and the dotted line is ruled out as the solution.
The weak solution being the solution of Eq.~\eqref{weak equation} with the RH condition is given by the solid line with a discontinuity at $\sigma^*$.
The discontinuity determined by the RH condition is called ``shock''.\footnote{
The terminology comes from the ``shock wave" in gas dynamics.
}
\begin{figure}
\begin{center}
\includegraphics[width=80mm]{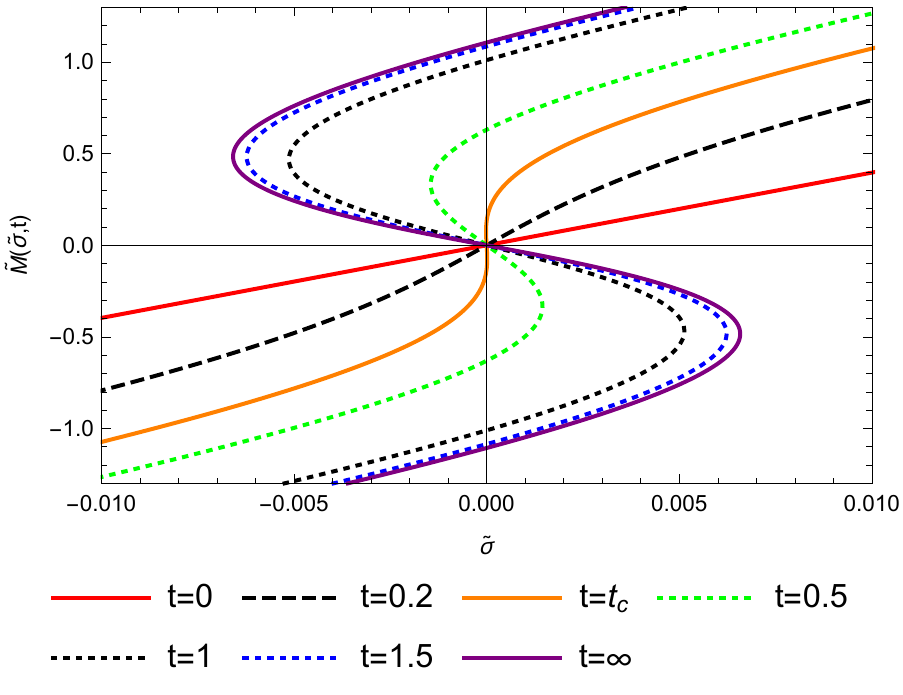}
\includegraphics[width=80mm]{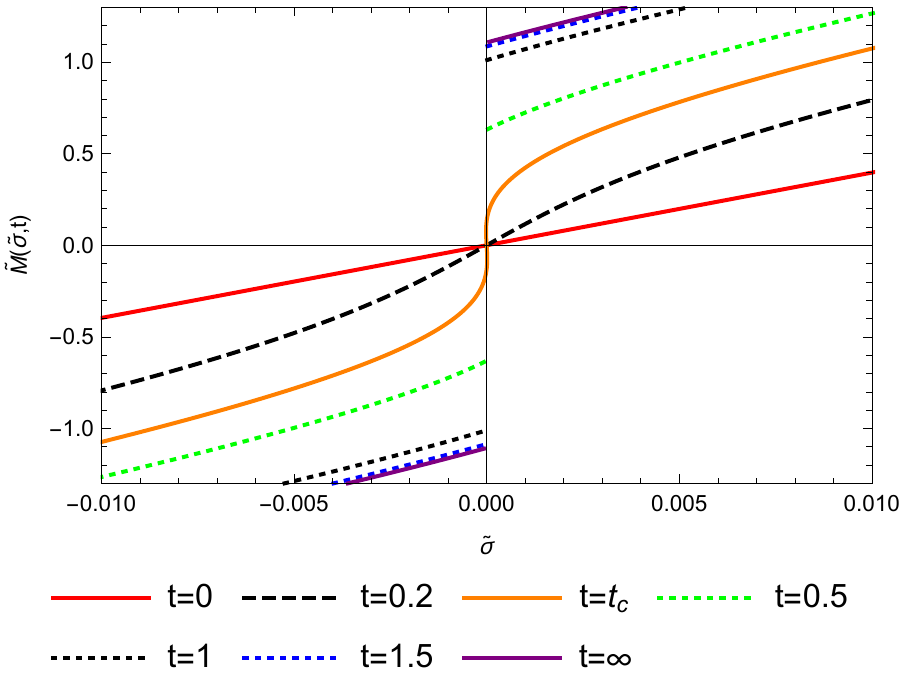}
\end{center}
\caption{The evolution of the mass function $M\fn{\sigma;t}$ at zero temperature and zero density. 
The initial condition is set to $g_0:=G_0\Lambda_0^2/2\pi^2=2.01$.
Left: the solution of Eq.~\eqref{caracta3}. Right: the weak solution.}
\label{weak_mass}
\end{figure}

\begin{figure}
\begin{center}
\includegraphics[width=80mm]{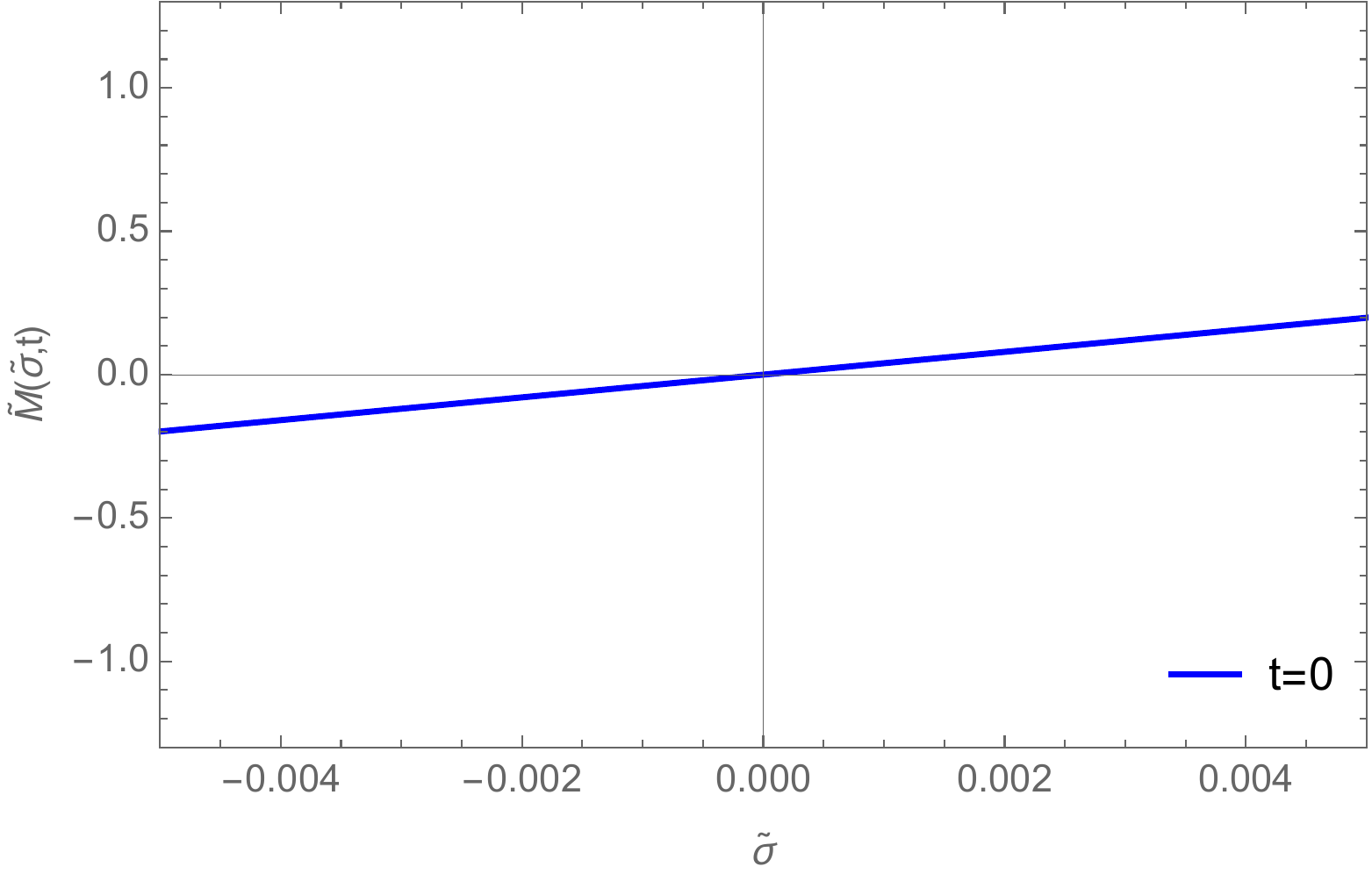}
\includegraphics[width=80mm]{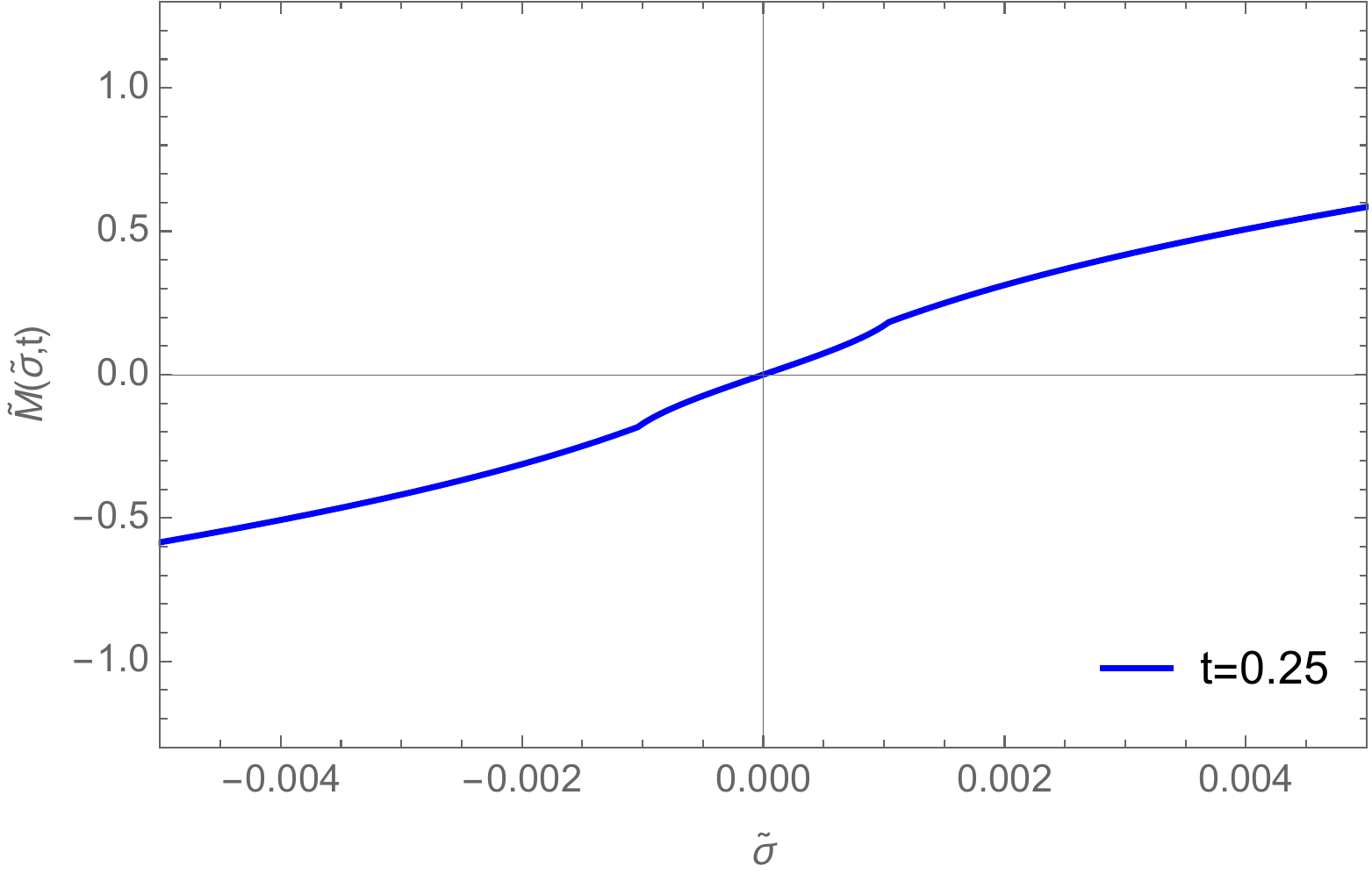}
\includegraphics[width=80mm]{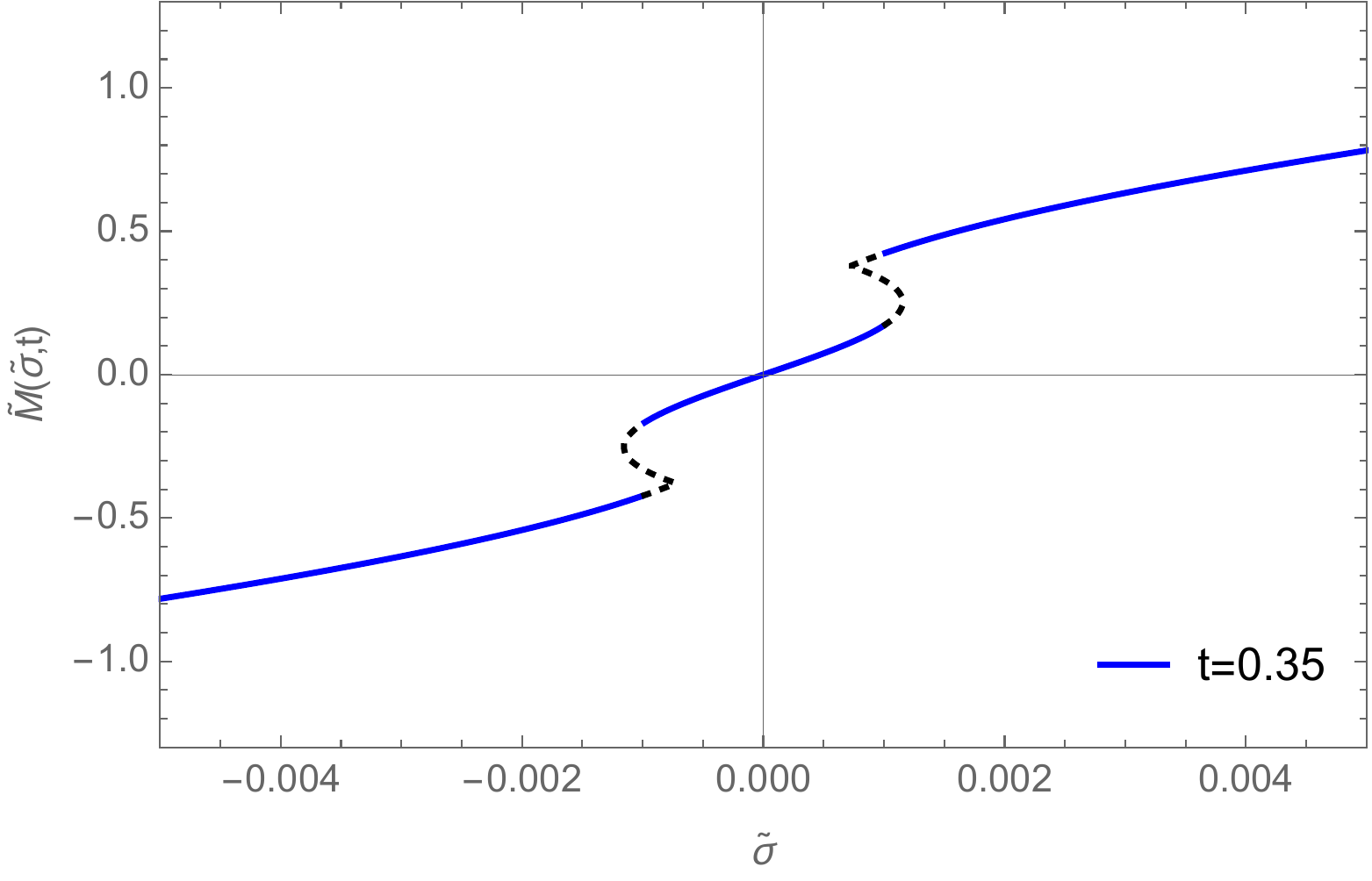}
\includegraphics[width=80mm]{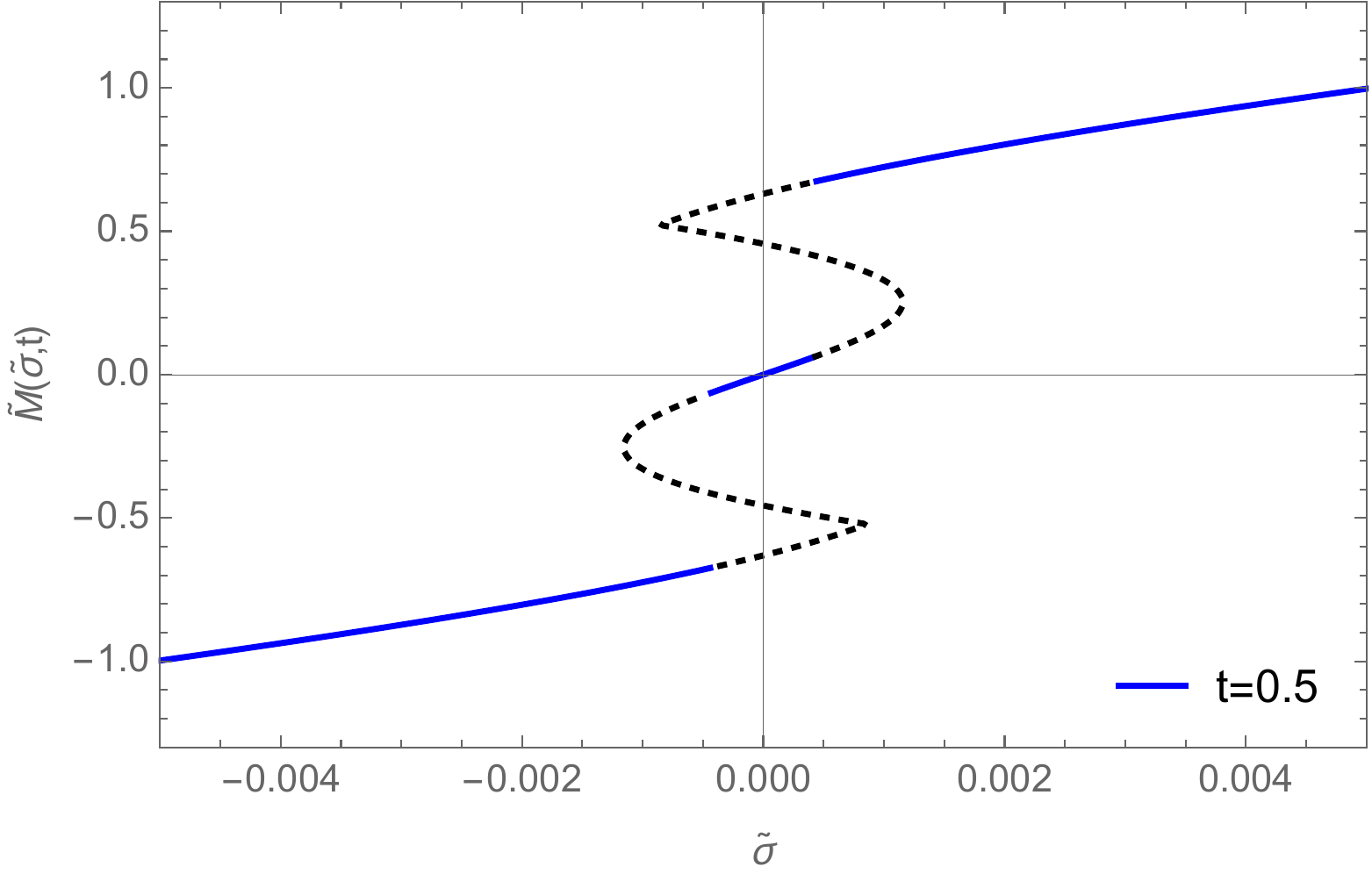}
\includegraphics[width=80mm]{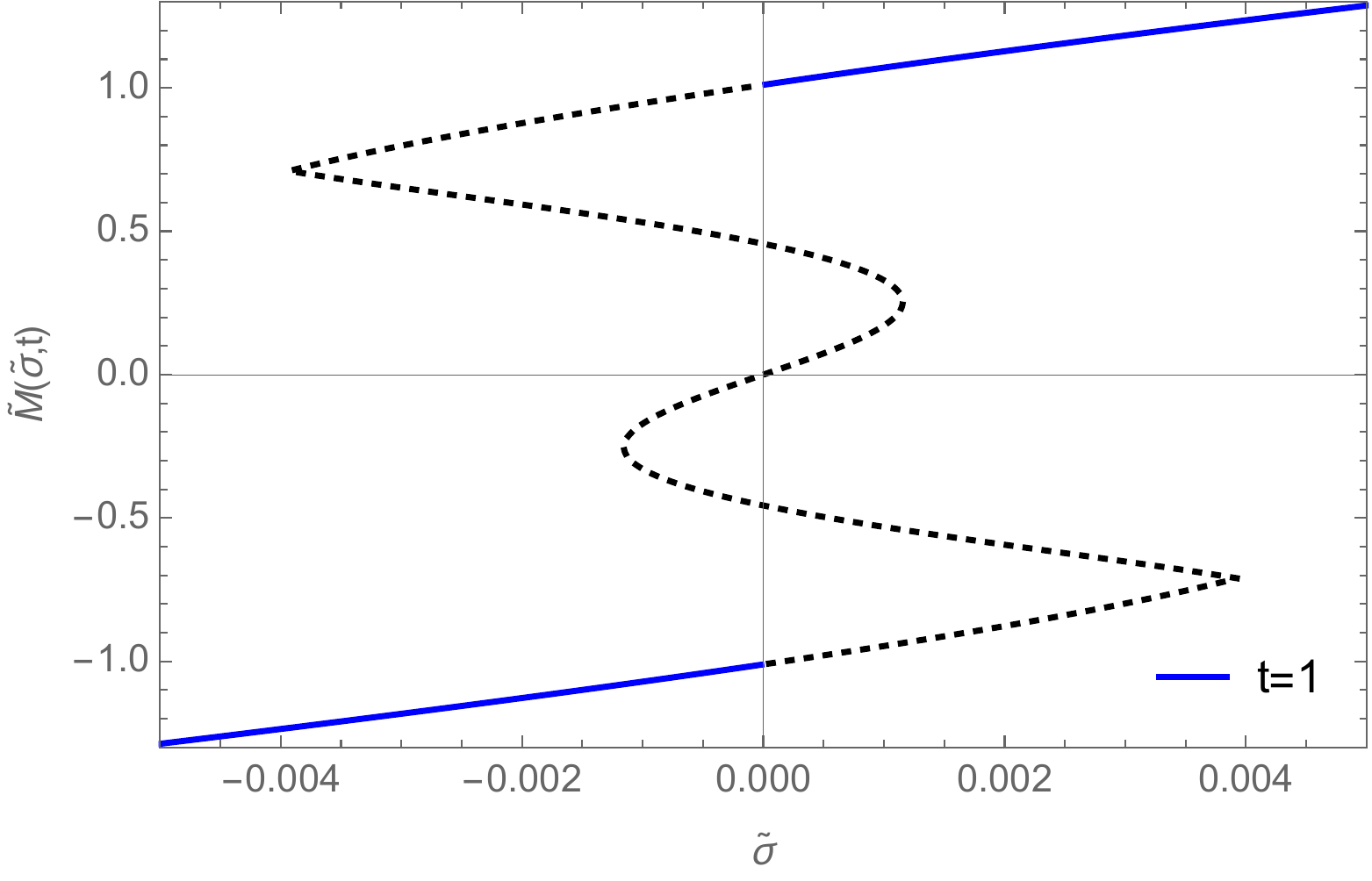}
\includegraphics[width=80mm]{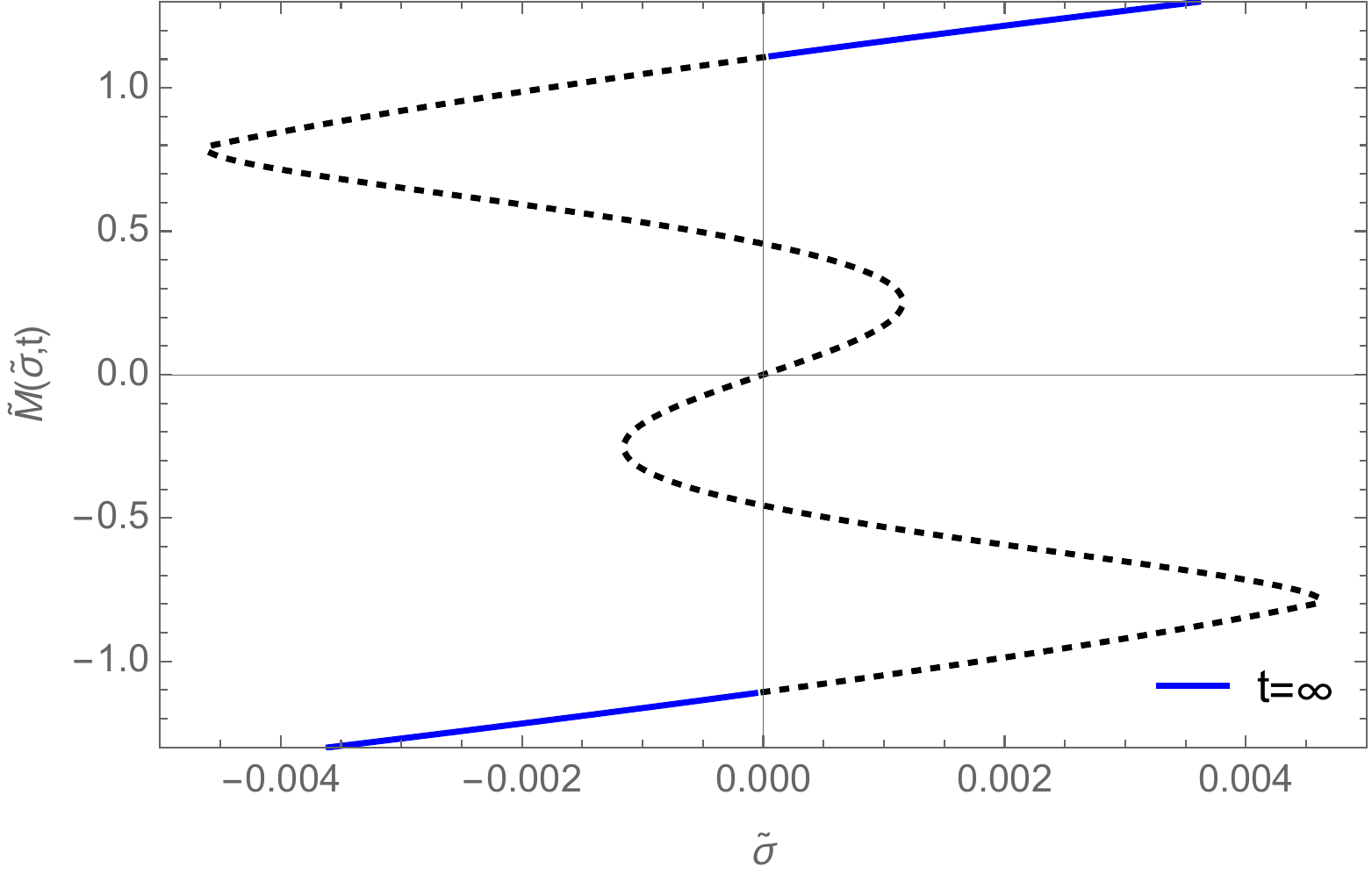}
\end{center}
\caption{The evolution of the mass function at zero temperature and finite density for $g_0:=G_0\Lambda_0^2/2\pi^2=2.01$. 
The blue line corresponds to the weak solution. The black dotted line is removed by the RH condition.}
\label{weak_mass_first}
\end{figure}

As an example, the evolution of the mass function at vanishing temperature and density is shown in Fig.~\ref{weak_mass}, where the dimensionless mass function $\tilde M=M/\Lambda_0$ and the dimensionless field $\tilde \sigma=\sigma/\Lambda_0$ are defined; see Eqs.~\eqref{dimlessbeta} and \eqref{dimlessvariables} in the section~\ref{numerical analysis}.
At the critical scale $t_c$, the slope of the mass function at $\tilde \sigma=0$ becomes infinity, which corresponds to the divergence of the four-fermi coupling constant.
After $t_c$, the solution of the mass function becomes the multi-valued function.
Applying the RH condition, the solution is uniquely determined; see the right-hand panel of Fig.~\ref{weak_mass}.
Next, let us consider the case at zero temperature and finite density.
The evolution of the mass function in this case is shown in Fig.~\ref{weak_mass_first}.
We see that two singularities arise at the non-zero field values, $\tilde\sigma\neq 0$. 
They move towards the origin $\tilde \sigma=0$ and eventually merge with lowering the scale $t$.
For such a behavior of the mass function, we observe the first-order phase transition as will be seen in the next section.

The mass function given as the weak solution satisfies
\al{
M\fn{\sigma\fn{t};t}=- M\fn{-\sigma\fn{t};t},
}
since it reflects the chiral transformation $\sigma \to -\sigma$.
Then, the dynamical mass is given by
\al{
\label{dynamical quark mass}
M_\text{phys}
=\lim _{\sigma\to 0_+}M\fn{\sigma;\infty}
=-\lim _{\sigma\to 0_-}M\fn{\sigma;\infty}.
}

To summarize so far, the multi-valued mass function as a solution of the characteristics \eqref{cha1}--\eqref{cha3} is uniquely determined for any field value of $\sigma$ by the RH condition.
Instead, the mass function with the RH condition has the discontinuities.
Such a solution is defined mathematically as the weak solution of Eq.~\eqref{weak equation} in which the mass function and its beta function are not differentiated by $t$ and $\sigma$.
The physical dynamical mass is given by Eq.~\eqref{dynamical quark mass}.

Finally, we comment on the Legendre effective potential and its convexity and concavity.
Let us start with defining the generating functional for the connected Green function,
\al{\label{wdefinition}
W\fn{j;t}:=V\fn{\sigma=0;t;j}.
}
Here, the initial condition for the fermionic potential is given by
\al{
V\fn{\sigma;t=0;j}=\frac{G_0}{2}\sigma^2+j\sigma,
}
where the last term is the source term.
From the Legendre transformation of Eq.~\eqref{wdefinition}, we can define the Legendre effective potential,
\al{
V_{\rm L}\fn{\phi;t}:=j \,\phi\fn{j;t}-W\fn{j;t}.
}
This potential is a function of the chiral condensate given by
\al{
\phi\fn{j;t}:=\frac{\p W\fn{j;t}}{\p j}
=\bra \bar \psi \psi \ket_{j}.
}
Note that the source $j$ is regarded as a function of $\phi$.
It is known that the Legendre effective potential $V_{\rm L}$ becomes the convex form~\cite{Fukuda:1975ib,O'Raifeartaigh:1986hi,Tetradis:1992qt,Litim:2006nn}.
The weak solution method automatically gives us the convex Legendre effective potential as the convex envelope of the nonconvex function obtained from the solution of characteristic equations \eqref{cha1}--\eqref{cha3}.
See discussions in \cite{Aoki:2014ola} on its proof.

\subsection{RG flow of four-Fermi coupling constant within weak solution method}
We discuss the RG flow of the four-Fermi coupling constant in view of the weak solution.
Since the mass function $M\fn{\sigma;t}$ is given by
\al{
M\fn{\sigma;t}=\p_\sigma V\fn{\sigma;t}=G_\Lambda \sigma +\cdots,
}
the behavior of the origin of $M\fn{\sigma;t}$ is described by the four-Fermi coupling constant.
At the critical scale $t_{\rm c}$ at which the four-Fermi coupling constant diverges, the slope of the mass function at $\sigma=0$ becomes infinity; see the orange line in Fig.~\ref{weak_mass}.
After the critical scale, it becomes negative whereas the RG flow of the four-Fermi coupling constant given by Eq.~\eqref{dimensionless rescaled four-fermi coupling constant} cannot be evaluated due to the divergence. 
We have seen in the previous section that in the weak solution method with the RH condition, the mass function with the discontinuities is defined mathematically and uniquely determined for any field value of $\sigma$.
Note that in the case where the first-order phase transition takes place, the slope of the mass function at the origin does not become infinity; see Fig.~\ref{weak_mass_first}.
That is, the RG flow of the four-Fermi coupling constant does not diverge at any scale.

Next, we discuss the relation between the inverse four-Fermi coupling constant $ g:=1/G_\Lambda$ and the Legendre effective potential $V_{\rm L}$.
We start with the well-known relation,
\al{
\label{WV_L}
-\frac{\p^2W}{\p j^2}\cdot \frac{\p ^2V_{\rm L}}{\p \phi^2}=1,
}
where $W\fn{j;t}$ is given in Eq.~\eqref{wdefinition}.
The second derivative of $W$ with respect to $j$ corresponds to the susceptibility:
\al{
\chi \fn{j} := \frac{\p \bra \bar \psi \psi \ket _{j}}{\p j} 
= -\frac{\p^2W}{\p j^2}
=\bra (\bar \psi \psi)^2 \ket_{j} -  (\bra \bar \psi \psi \ket_{j})^2.
}
Together with Eq.~\eqref{WV_L}, we obtain
\al{
\frac{\p ^2V_{\rm L}\fn{\phi;t}}{\p \phi^2} = \frac{1}{\chi \fn{j}}.
}
When the second-order phase transition occurs, the susceptibility diverges: $\chi\to \infty$.
That is, the left-hand side $\frac{\p ^2V_{\rm L}\fn{\phi;t}}{\p \phi^2}$ vanishes at the critical scale $t=t_{\rm c}$.
More explicitly, we see
\al{
\frac{\p ^2V_{\rm L}\fn{\phi;t}}{\p \phi^2}
=m^2 +\lambda \phi^2 +\cdots.
}
At the origin of $V_{\rm L}$, the mass-squared $m^2$, i.e. the curvature of $V_{\rm L}$, corresponds to the inverse susceptibility $1/\chi$.
The divergence of $\chi$ means the vanishing mass-squared $m^2=0$.
As mentioned in the section \ref{njl model and weak nprg}, the four-Fermi coupling constant corresponds to the chiral susceptibility:
\al{
G_\Lambda= \frac{1}{2}\left.\frac{\p^2 W\fn{j}}{\p j^2}\right|_{j=0}+G_{\Lambda_0}.
}
Then, the divergence of $G_\Lambda$ is the signal of the second-order phase transition, and we see that the inverse four-Fermi coupling constant corresponds to the curvature at the origin of the Legendre effective potential:
\al{\label{m2eqtildeg}
m^2 = \frac{1}{G_\Lambda}.
}
The RG flow of the dimensionless rescaled inverse four-Fermi coupling constant $\tilde \kappa=\tilde G^{-1}=2\pi^2/(G_\Lambda \Lambda^2)$ is given by
\al{
\tilde \kappa\fn{t} =1 -(1-\tilde \kappa_0)\e ^{2t},
}
where $\tilde \kappa_0$ is the initial value of $\tilde \kappa$ at the initial scale $t=0$.
For $\tilde \kappa_0 <1$, $\kappa\fn{t}$ becomes negative after $t=t_{\rm c}$ at which $\tilde \kappa\fn{t_{\rm c}}=0$.
Thus, from the relation \eqref{m2eqtildeg}, one can see that the negative $\tilde \kappa$ corresponds to  the negative curvature at the origin of the Legendre effective action.
The negative curvature at the origin means that the non-vanishing vacuum $\bra \phi\ket = \bra \bar \psi \psi \ket\neq 0$ exists.
However, one cannot evaluate the value of the chiral condensate by only observing the divergence of the four-Fermi coupling constant in the second-order phase transition.
Besides, the first-order phase transition is not captured by only looking for the behavior of the four-Fermi coupling constant.

In the next section, we analyze the weak renormalization group at finite temperature and density and show the chiral phase structures of the NJL model.
\section{Numerical analysis}\label{numerical analysis}
In this section, we show the chiral phase transitions within the NJL model described by Eq.~\eqref{dimensionful RG equation}.
To this end, we first give the RG equation written by the dimensionless variables.
Then, the evolutions of the mass function with second- and first-order phase transitions are shown for a benchmark value of the four-Fermi coupling constant.
The chiral phase diagram of the NJL model is shown.
In the last subsection, we show our results at a benchmark point in comparison with results of other works.

\subsection{Dimensionless RG equations and initial conditions}
We make the dimensionless beta function of $F\fn{\sigma;t}$ given in Eq.~\eqref{dimensionful RG equation} at finite temperature and density: 
\al{
\tilde F\fn{\tilde \sigma;t}=-\frac{\e^{-3t}}{\pi^2}
\bigg[
{\tilde E} +{\tilde T}\log \fn{1+\e^{-{\tilde \beta} ({\tilde E}-{\tilde \mu})}} +{\tilde T} \log \fn{1+\e^{-{\tilde \beta}({\tilde E}+{\tilde \mu})}}
\bigg],
\label{dimlessbeta}
}
where we defined the dimensionless variables,
\al{
\tilde E&=\frac{E}{\Lambda_0}=\sqrt{\e^{-2t}+\tilde M^2},&
\tilde M&=\frac{M}{\Lambda_0},&
\tilde \sigma&=\frac{\sigma}{\Lambda_0^3},&
\tilde\beta^{-1}=\tilde T&=\frac{T}{\Lambda_0},&
\tilde \mu&=\frac{\mu}{\Lambda_0}.&
\label{dimlessvariables}
}
One obtains evolutions of the dimensionless mass function and the dimensionless effective potential by solving the dimensionless characteristic equations,
\al{
&\frac{\df \tilde \sigma \fn{t}}{\df t}=\frac{\p \tilde F}{\p \tilde M},&
&\frac{\df \tilde M\fn{\tilde \sigma ,t}}{\df t}=0,&
&\frac{\df \tilde V\fn{t}}{\df t}=\tilde M\frac{\p \tilde F}{\p \tilde M}- \tilde F,
\label{dimlesscaracta}&
}
with the initial conditions at $t=0$ ($\Lambda=\Lambda_0$), 
\al{
{\tilde V}\fn{\tilde\sigma;t=0}&= \pi^2g_0 {\tilde \sigma^2},&
{\tilde M}\fn{\tilde\sigma;t=0}&=2\pi^2 g_0 {\tilde \sigma},&
}
where $g_0:=G_0\Lambda_0^2/2\pi^2$.
Note that the critical value of $g_0$ at vanishing temperature and density is $g_0{}_\text{c}=1$.
Hence, for $g_0>1$, the D$\chi$SB is observed.
 
\subsection{Phase transitions}
\begin{figure}
\begin{center}
\includegraphics[width=150mm]{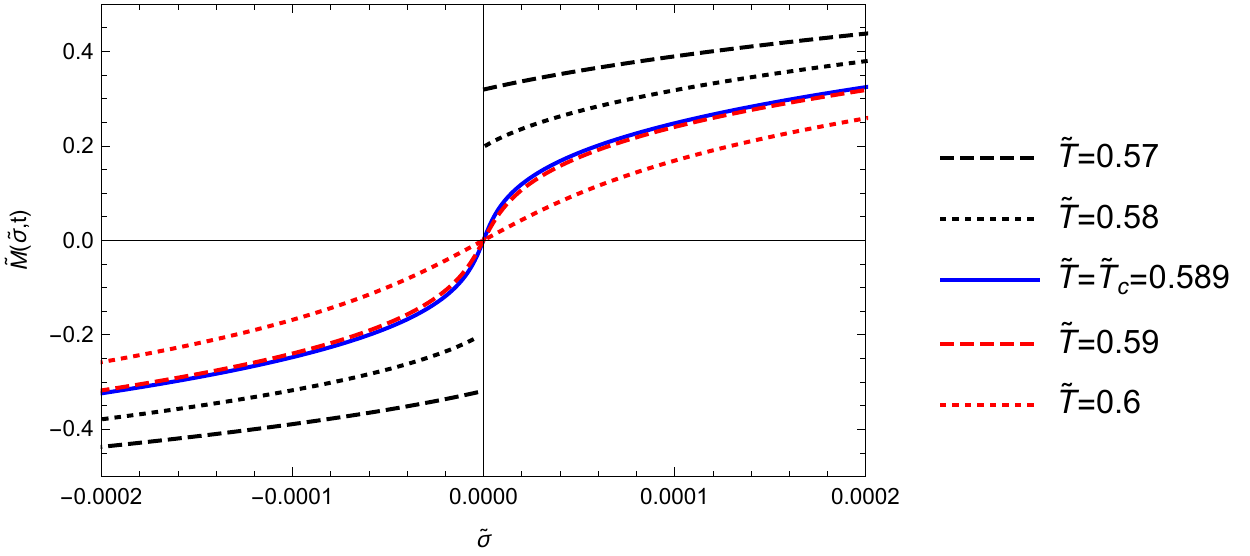}
\end{center}
\caption{The mass function around the second-order phase transition at finite temperature and zero density in the IR limit $t\to \infty$ ($\Lambda\to 0$).
The initial value of the four-Fermi coupling constant is set to $g_0=2$.
}
\label{second_phase_trans}
\end{figure}
\begin{figure}
\begin{center}
\includegraphics[width=80mm]{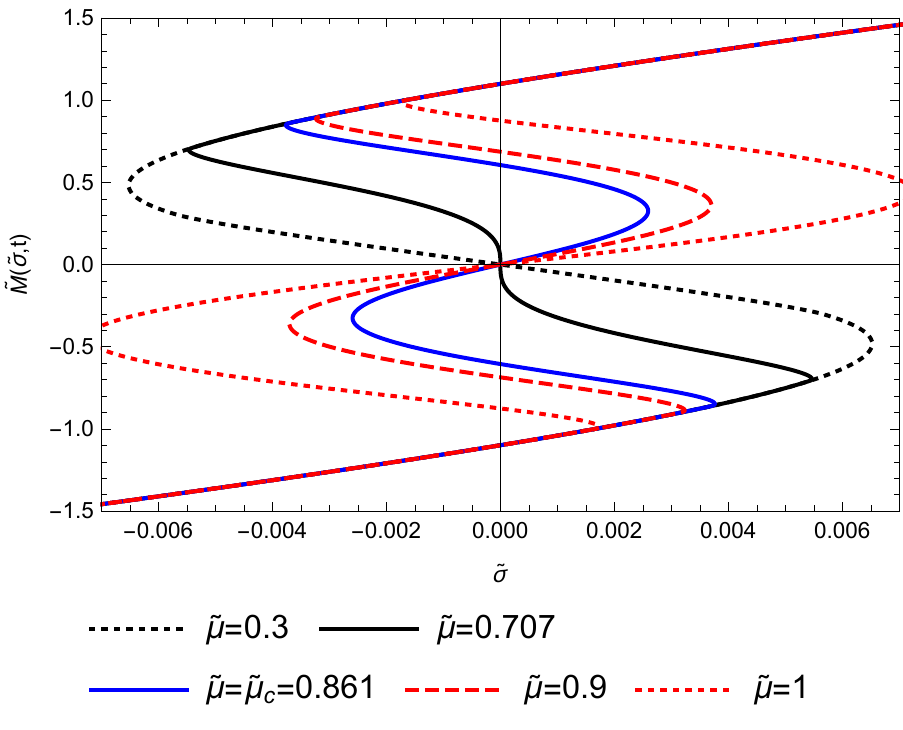}
\includegraphics[width=80mm]{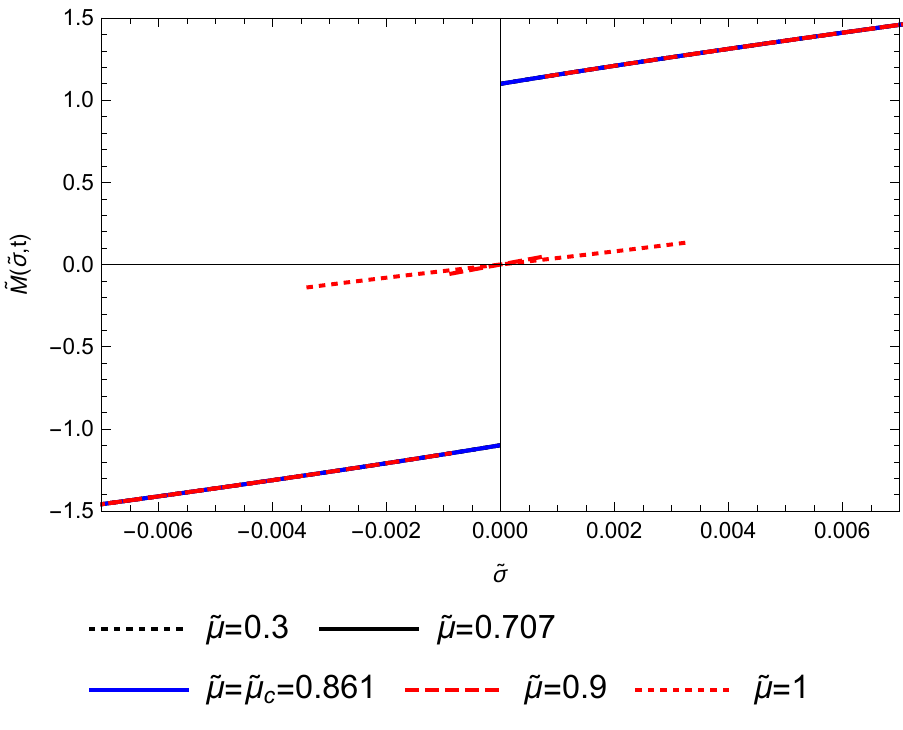}
\end{center}
\caption{The mass function around the first-order phase transition at zero temperature and finite density in the IR limit $t\to \infty$ ($\Lambda\to 0$).
Left: The solutions of Eq.~\eqref{dimlesscaracta}. 
Right: The weak solution of the mass function.
The initial value of the four-Fermi coupling constant is set to $g_0=2$.
}
\label{first_phase_trans}
\end{figure}
Here setting $g_0=2$ as a benchmark point, we show the dependence of the mass function on temperature and density.
First, let us consider the zero density case $\mu=0$.
We show the dependence of the mass function in the IR limit $t\to \infty$ ($\Lambda\to 0$) on temperature in Fig.~\ref{second_phase_trans}.
With increasing $\tilde T$, the value of the mass function at $\tilde \sigma\to 0$ smoothly becomes smaller.
This behavior corresponds to the second-order phase transition.
Note that the dynamical mass is given in Eq.~\eqref{dynamical quark mass}.
The mass function at $\tilde \sigma\to 0$ vanishes at ${\tilde T}_\text{c}\simeq 0.589$ which is the critical temperature.
Next, we consider the finite chemical potential and vanishing temperature case.
In this case, the evolution of the mass function in the IR limit  is shown in Fig.~\ref{first_phase_trans}.
We see that the slope of the origin of the mass function changes from negative to positive at $\tilde \mu\simeq 0.707$; see the left-hand side of Fig.~\ref{first_phase_trans}.
This means that the four-Fermi coupling constant becomes positive via the divergence, and thus the Legendre effective potential has a local minimum at its origin.
Here we call this behavior the ``second-order" although it does not always correspond to the phase transition.
Note that there is still a non-trivial global minimum at $\tilde \mu\simeq 0.707$.
Above $\tilde \mu_\text{c}\simeq 0.861$, the discontinuities of the mass function in the IR limit are located at non-vanishing $\sigma$.
Then, when taking the limit $\sigma\to0$, the physical dynamical mass suddenly vanishes with increasing $\tilde \mu$, i.e., $M\fn{\sigma\to 0;t\to \infty;T=0,\mu<\mu_\text{c}}\neq 0 \to M\fn{\sigma\to 0;t\to \infty;T=0,\mu>\mu_\text{c}}=0$.
This phase transition is of first-order.
We see that the weak solution method can capture the phase transitions with both second- and first-order.
In the next subsection, we show the phase diagram on the $g_0$--$\tilde \mu$--$\tilde T$ plane.

\subsection{Phase diagram}

\begin{figure}
\begin{center}
\includegraphics[width=80mm]{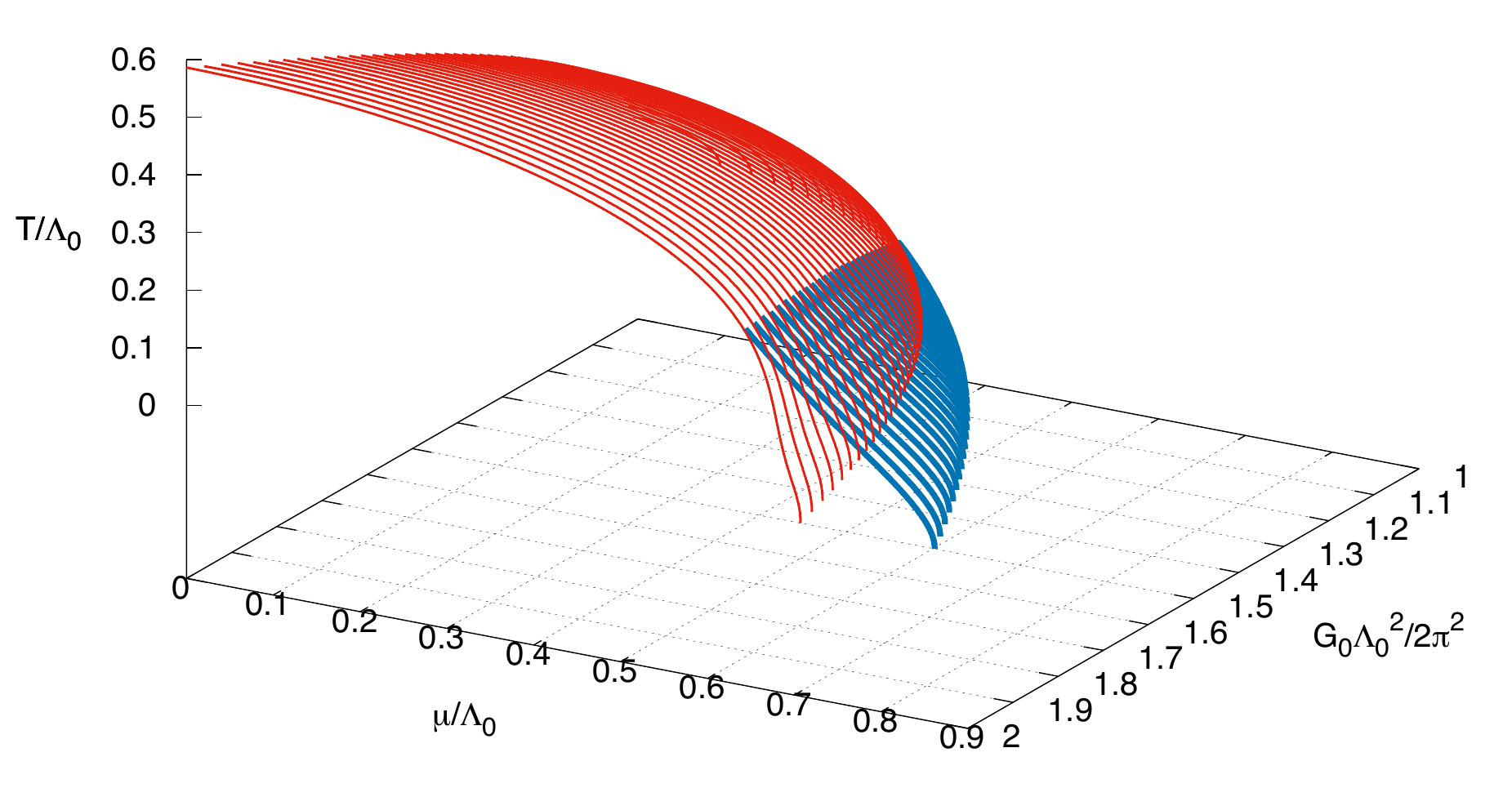}
\includegraphics[width=80mm]{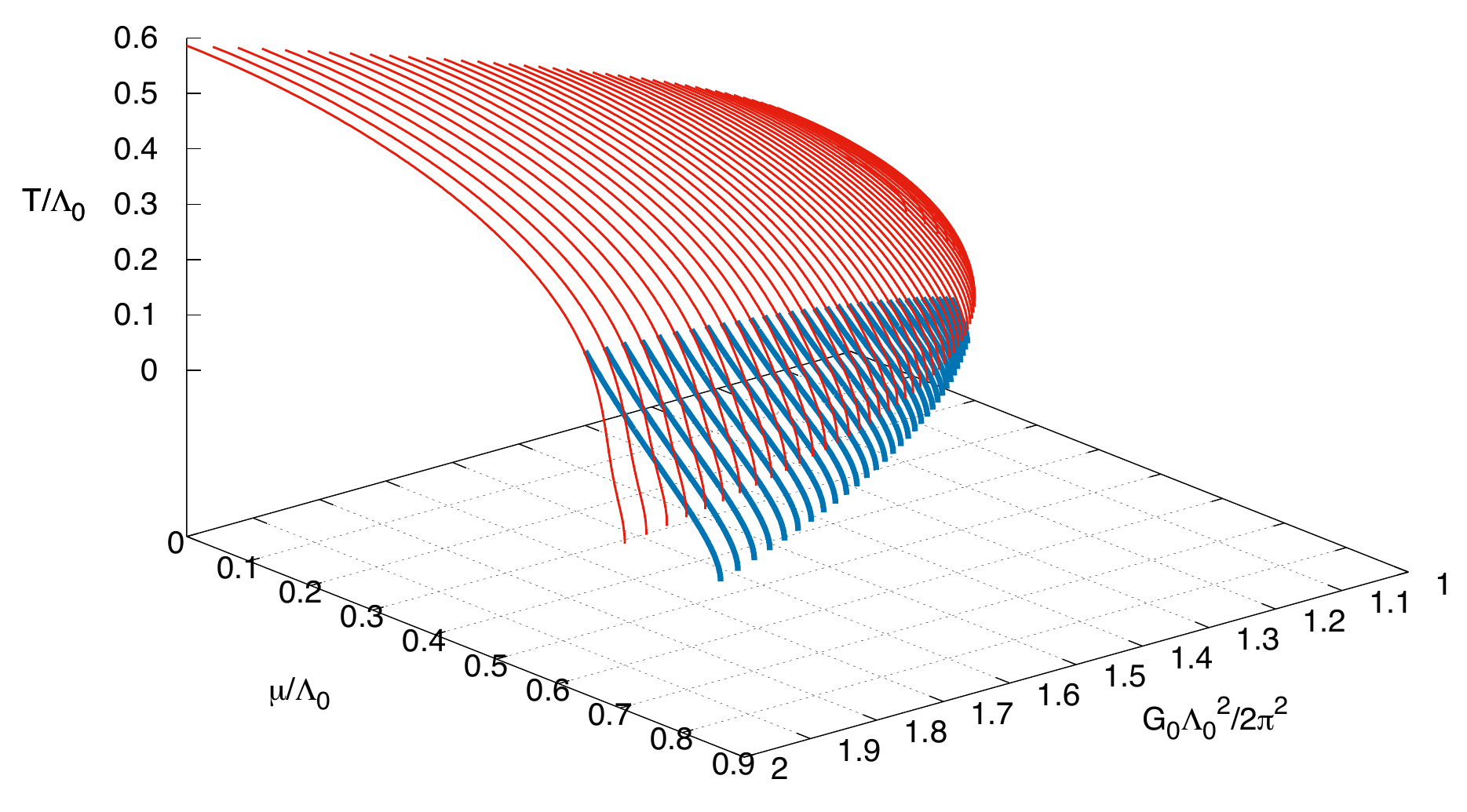}
\includegraphics[width=80mm]{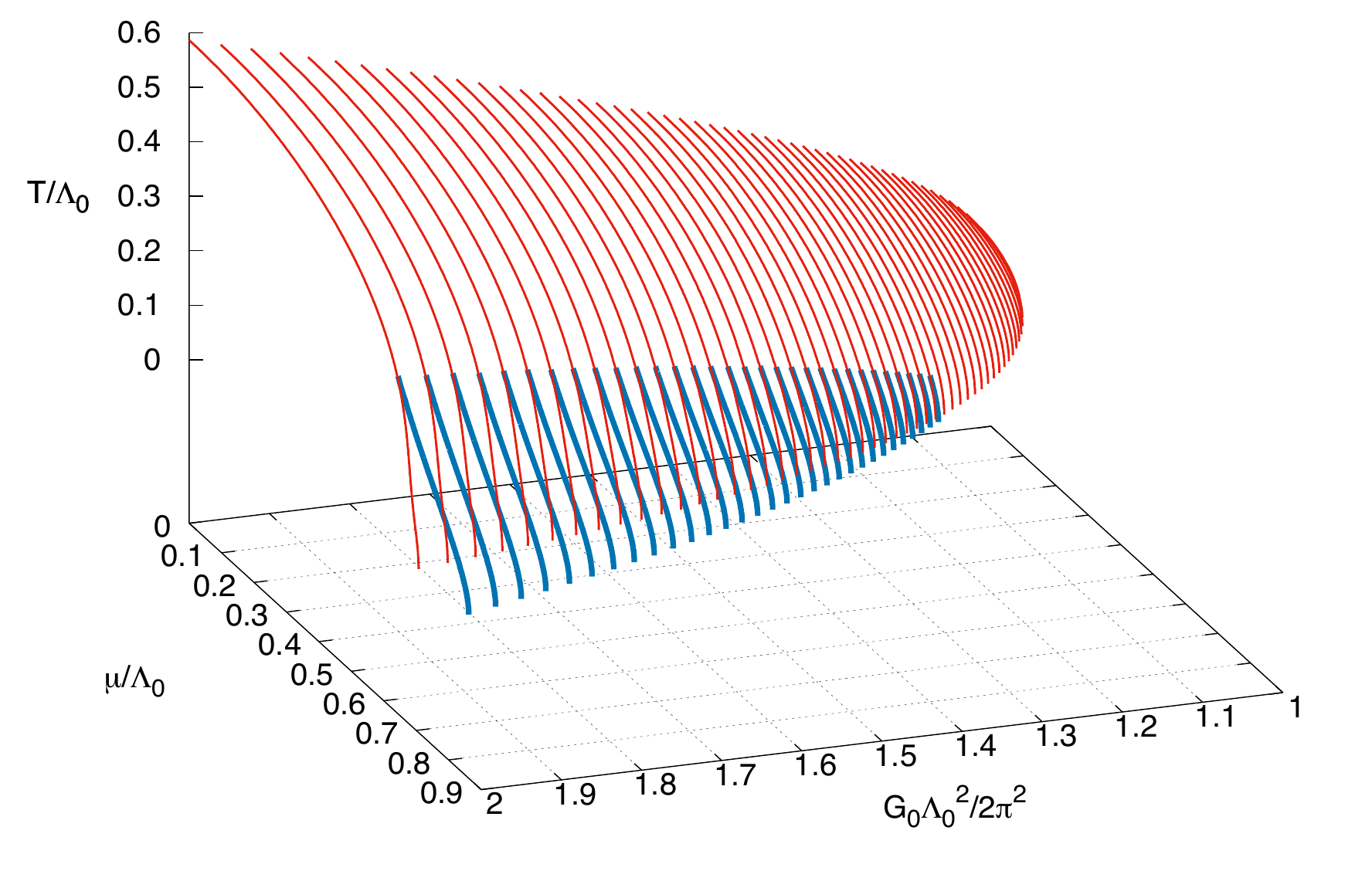}
\includegraphics[width=80mm]{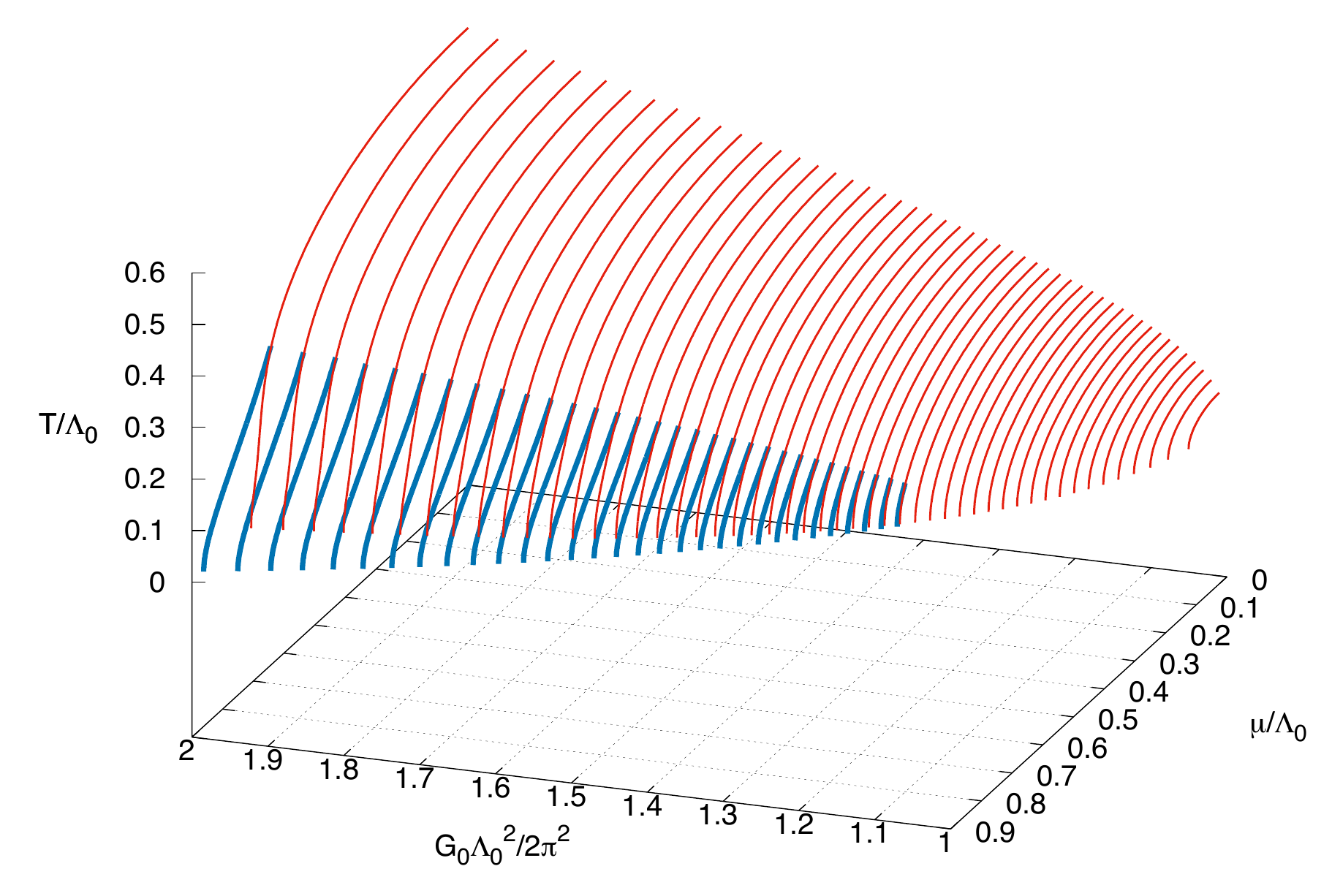}
\includegraphics[width=80mm]{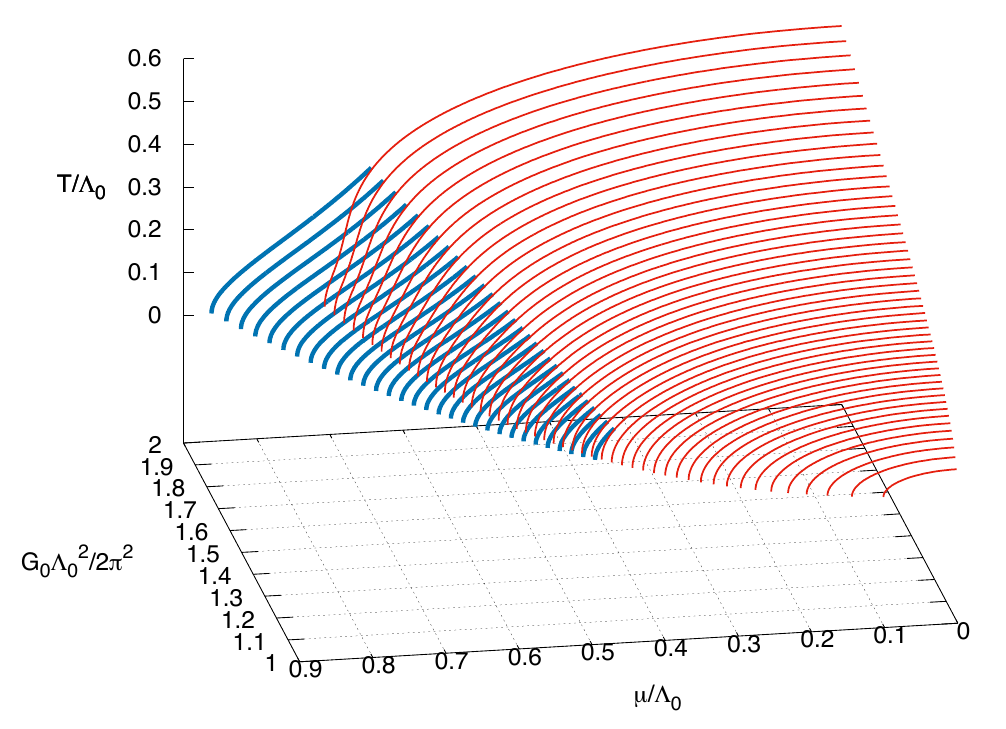}
\includegraphics[width=80mm]{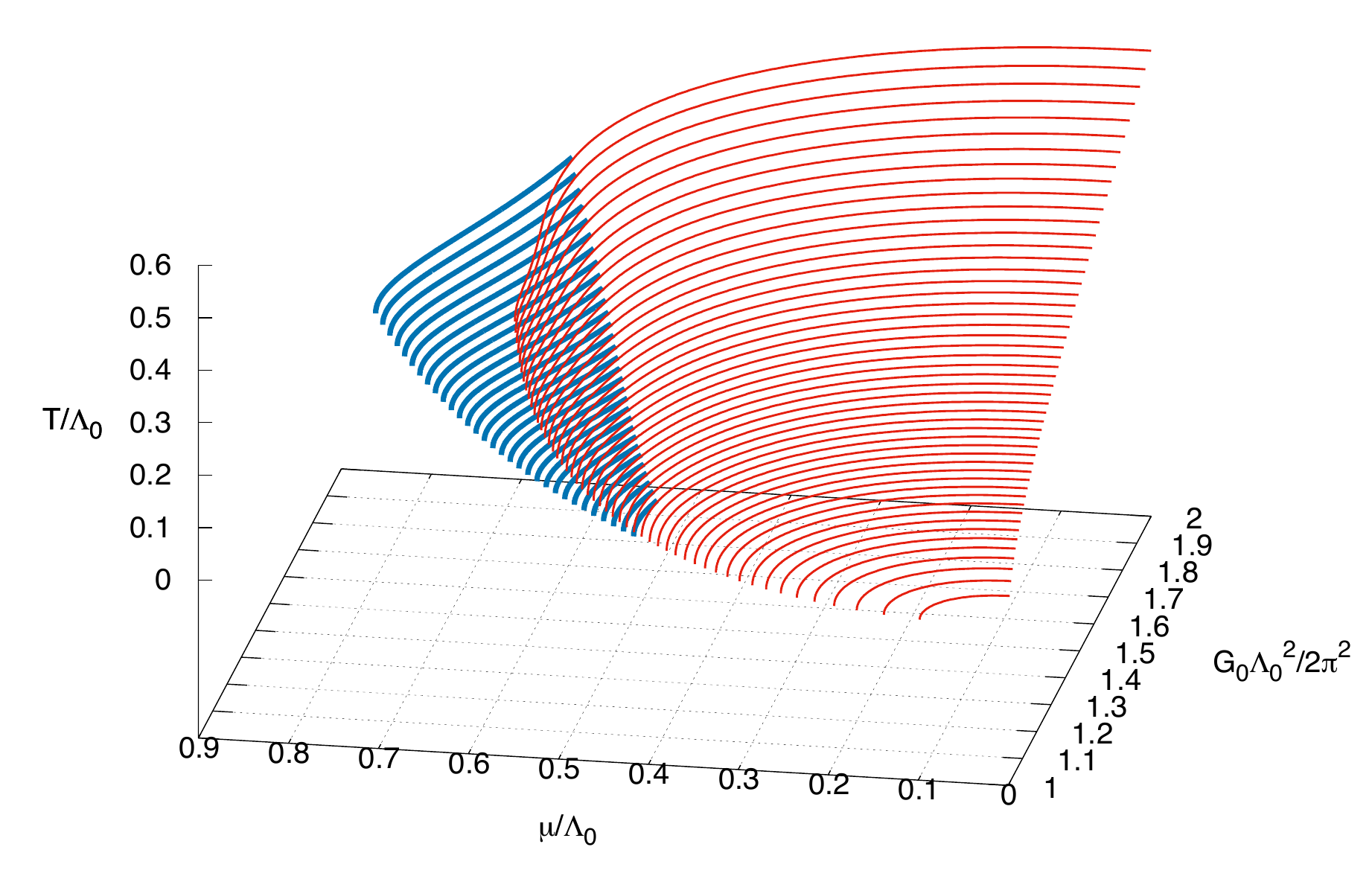}
\end{center}
\caption{The chiral phase diagram on the $g$--$\tilde \mu$--$\tilde T$ plane.
The biggest and smallest phase boundaries correspond to the cases of $g_0=2$ and $g_0=1.0101$, respectively.
}
\label{3dphase}
\end{figure}
\begin{figure}
\begin{center}
\includegraphics[width=100mm]{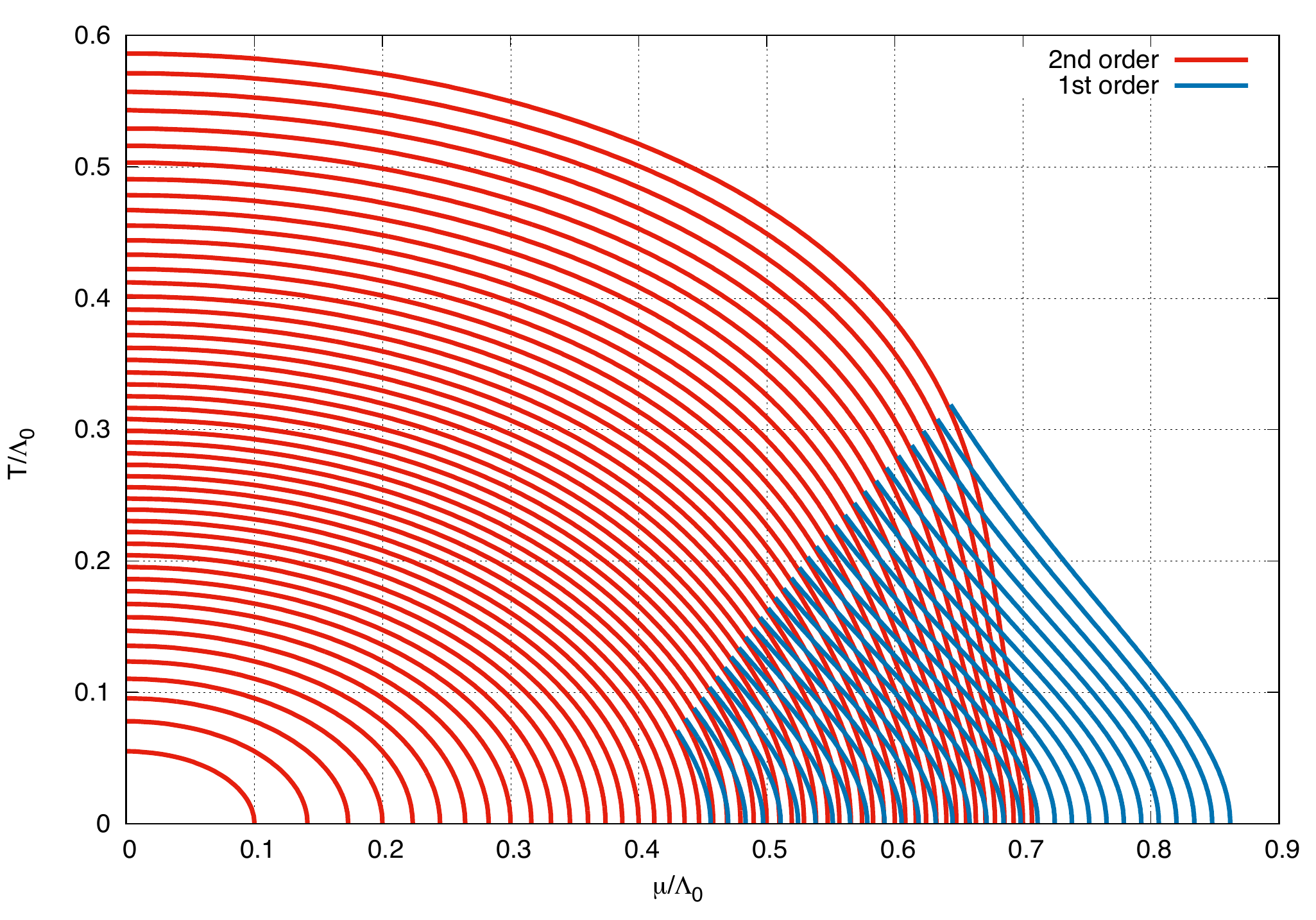}
\end{center}
\caption{The chiral phase diagram on the $\tilde \mu$--$\tilde T$ plane.
The biggest and smallest phase boundaries correspond to the cases of $g_0=2$ and $g_0=1.0101$, respectively.
}
\label{2dphase}
\end{figure}
\begin{figure}
\begin{center}
\includegraphics[width=100mm]{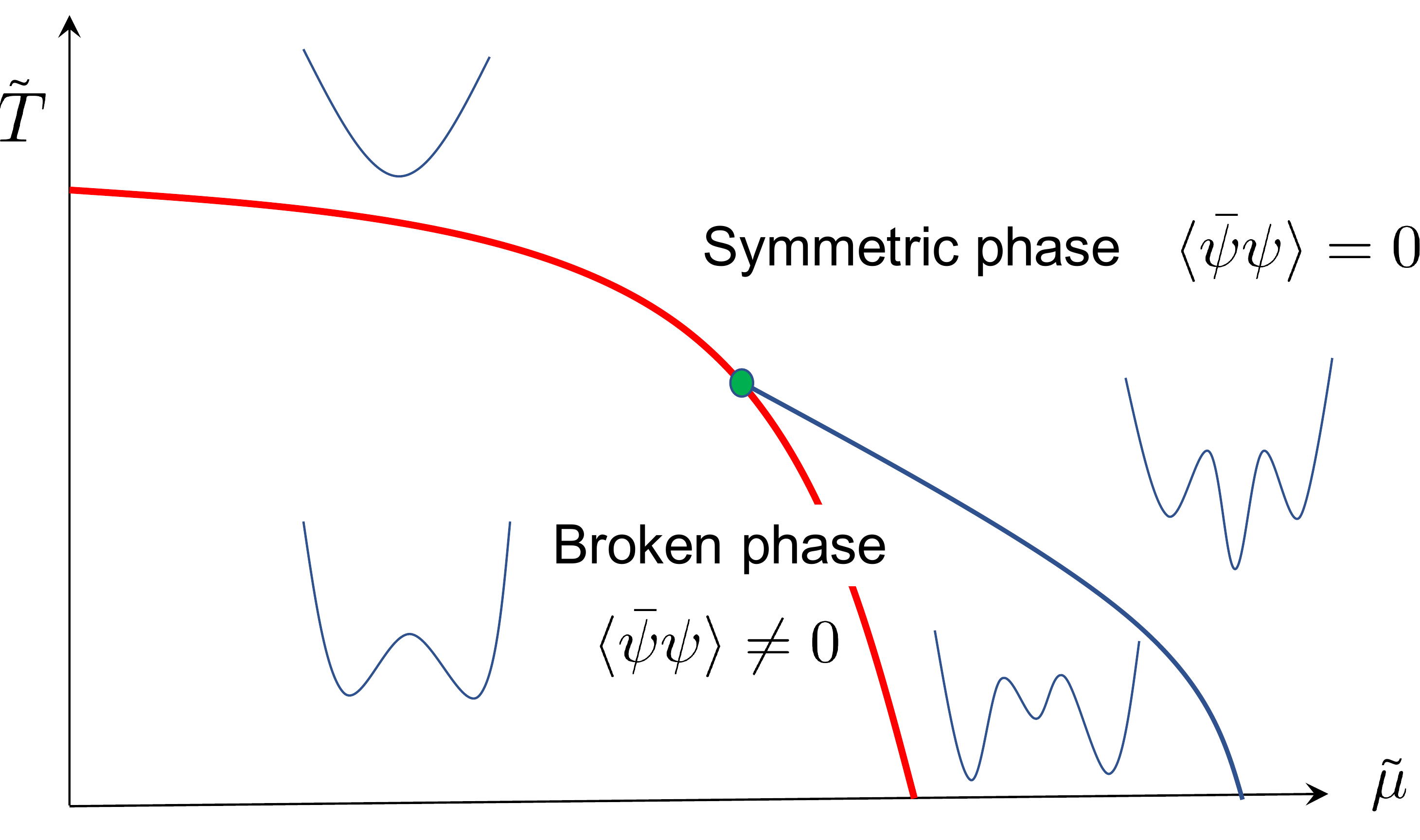}
\end{center}
\caption{Schematic figure of the chiral phase diagram on the $\tilde \mu$--$\tilde T$ plane with a fixed four-Fermi coupling constant and the shape of the bosonic effective potential.}
\label{2dphaseschematic}
\end{figure}
\begin{figure}
\begin{center}
\includegraphics[width=100mm]{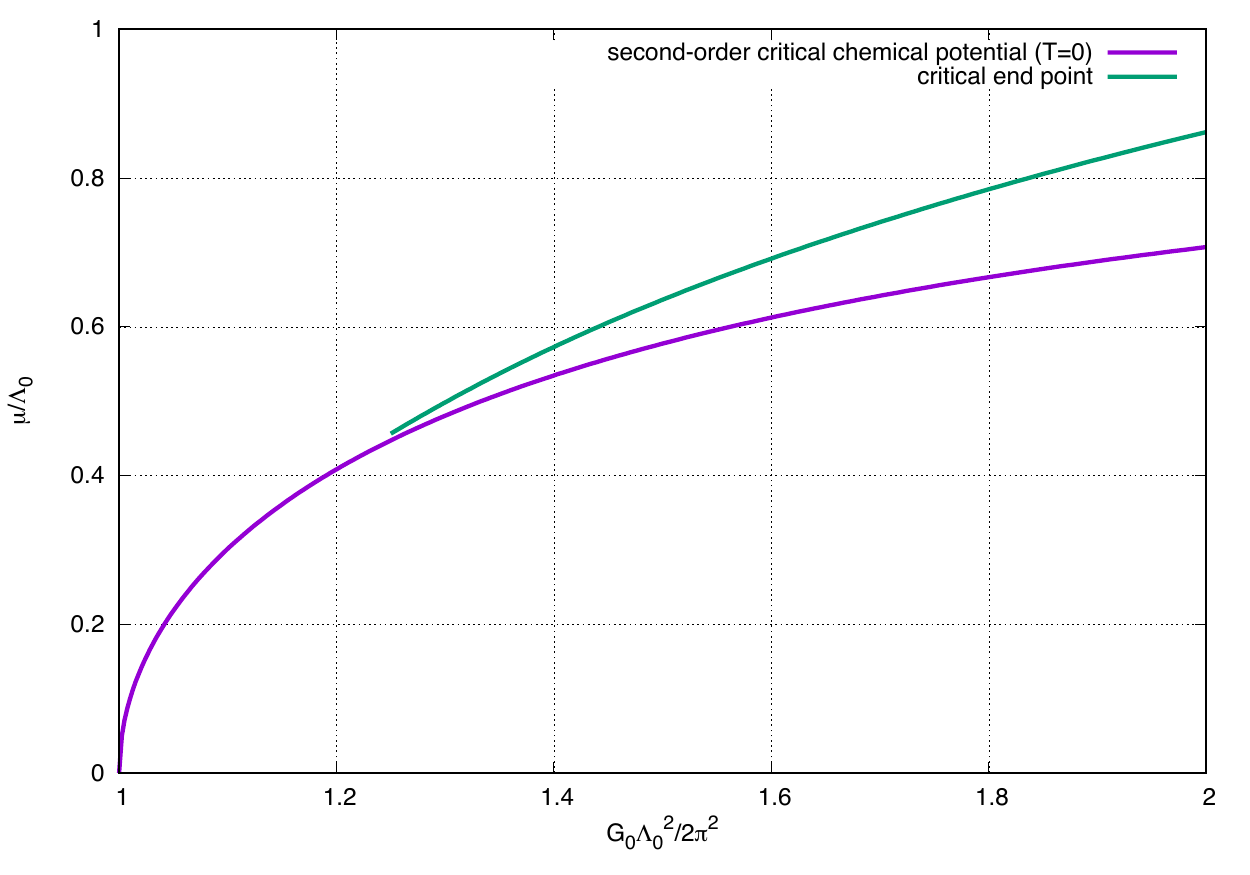}
\end{center}
\caption{The behaviors of the critical end-point and the second-order phase boundary at zero temperature with varying the four-Fermi coupling constant.}
\label{critical}
\end{figure}
In Fig.~\ref{3dphase}, the chiral phase diagram on the $g_0$--$\tilde \mu$--$\tilde T$ plane are shown.
The red line stands for the second-order phase boundary. 
As discussed in the section~\ref{njl model and weak nprg}, this boundary can be obtained by only the behavior of the four-Fermi coupling constant.
The blue line corresponds to the first-order phase boundary.
The biggest and smallest phase boundaries correspond to the cases of $g_0=2$ and $g_0=1.0101$, respectively.
We show the phase diagram on the $\tilde \mu$--$\tilde T$ plane in Fig.~\ref{2dphase}, which is obtained by the projection from the three dimensional phase diagram of Fig.~\ref{3dphase}.
A schematic figure of a phase boundary with a fixed four-Fermi coupling constant is shown in Fig.~\ref{2dphaseschematic}.
Note that the region, which is surrounded by the second- and first-order phase boundaries towards the high density side, is the broken phase.
As discussed in the previous subsection, although the four-Fermi coupling constant diverges at the second-order phase boundary, that is, the curvature of the origin of the effective potential becomes positive, the global minimum is still located at the non-vanishing chiral condensate.
In Fig.~\ref{critical}, the behaviors of the critical end-point (CEP) and the second-order phase boundary at $T=0$ with varying the four-Fermi coupling constant are shown.
At a finite value of the four-Fermi coupling constant $g_0\simeq 1.2$, they merge and the CEP vanishes.\footnote{
We see in Fig.~\ref{2dphase} that the first-order phase boundary abruptly disappears at a certain four-Fermi coupling constant.
However, the CEP should completely merge with the second-order phase boundary at $T=0$.
In order to see this, we need very precise numerical calculations.
Here, extrapolating the CEP towards the horizontal axis, we obtain the value of the four-Fermi coupling constant at which the critical end-point merges with the second-order phase boundary at $T=0$.
}
The first-order phase boundary vanishes for the smaller four-Fermi coupling constant than $g_0\simeq 1.2$.
We again emphasize that the chiral phase diagrams shown in this work are obtained by the RG equation without introducing the auxiliary field.

Here, we note the case where the finite bare mass is taken into account.
In this case, we could observe the crossover for smaller chemical potential rather than the second-order phase transition.
This behavior within the weak renormalization group is shown in \cite{Aoki:2014ola}.

\subsection{Comparisons with other models and methods}
Let us compare our results with other models and methods.
We first note that as shown in Refs.\,\cite{Aoki:1999dw,Terao:2000ae}, the FRG analysis in the large-$N$ limit is equivalent to the SDE with the ladder approximation.
The ladder approximated SDE is also equivalent to the MFA.
Therefore, these methods yield quantitatively the same results.
\tabcolsep = 7.5pt
\begin{table}
  \begin{center}
    \begin{tabular}{llcccc} \hline
      Model & Method & pion mass & $T_c$($\mu=0$)& CEP $(T_c, \mu_c) $ & $\kappa_\mu$ \\[-8pt]
      & & (or bare quark mass) & [MeV] & [MeV]  & \\ \hline\hline 
      This work & FRG (LPA) & chiral limit & $182$ & $(63, 278)$ & $1.900$ \\
      NJL (I)~\cite{Buballa:2003qv} & MFA & chiral limit & $222$ & $(112,286)$ & $1.974$ \\
      NJL (II)~\cite{Buballa:2003qv} & MFA & $135$\,MeV & --- & $(81, 330)$ & $1.974$\\
      NJL (III)& MFA & $138$\,MeV & $210$ & (48, 328) & $1.947$ \\
      QM~\cite{Schaefer:2004en} & FRG (LPA) & chiral limit & $143$ & $(52, 251)$ & $1.135$ \\
      QM~\cite{Braun:2011iz} & FRG (LPA) & $138$\,MeV & --- & --- & 1.375\\
      HQM~\cite{Pawlowski:2014zaa} & FRG & $138$\,MeV & $166$ & $(50, 291)$ & $1.397$ \\ 
      PQM (I)~\cite{Schaefer:2007pw} & MFA & $138$\,MeV & $184$ & $(150,168)$ & --- \\ 
      PQM (II)~\cite{Schaefer:2007pw} & MFA & $138$\,MeV & $184$ & $(163,164)$ & --- \\ 
      PQM (I)~\cite{Herbst:2010rf} & FRG & $138$\,MeV & $190$ & $(23, 292)$ & --- \\ 
      PQM (II)~\cite{Herbst:2010rf} & FRG & $138$\,MeV & $190$ & $(32, 293)$ & --- \\ 
      QCD~\cite{deForcrand:2002hgr} & Lattice ($i\mu$) & $m_qa=0.025$ & $174$ & --- & 0.498 \\
      QCD~\cite{Wu:2006su} & Lattice ($i\mu$) & $0.899\leq \frac{m_\pi}{m_\rho}\leq 0.943$ & $171$ & --- & 0.641 \\
      QCD~\cite{Allton:2003vx} & Lattice (Taylor) & $m_qa =0.1$ & $170$ & --- & 0.691 \\    \hline
    \end{tabular}
        \caption{The (pseudo-)critical temperature at vanishing chemical potential, the CEP and the curvature of the chiral phase boundary \eqref{curvature of PB} obtained from several models and methods in the two-flavor and three-color case. The abbreviation ``QM" stands for the quark-meson model. ``HQM" and ``PQM" are the QM with higher order quark-mesonic scattering processes and the Polyakov-QM models, respectively. The numbers (I), (II) in the PQM denote different values of the parameter $\hat\gamma$ governing the curvature of the critical Polyakov-loop temperature $T_0\fn{\mu}$; see \cite{Schaefer:2007pw,Herbst:2010rf}.
        }
            \label{tab:comparison}
  \end{center}
\end{table}

We compare the present work with the other ones in the two-flavor and three-color case ($N_\text{f}=2$, $N_\text{c}=3$).\footnote{
In the present setup, there are no significant differences between one-flavor and two-flavor cases. See the comments in the footnote \ref{comment on flavor case}.
}
For the mean-field analysis of the NJL model, we employ the following Lagrangian,
\al{
{\mathcal L}_\text{2fNJL}=\bar \psi (\Slash \p -\mu \gamma^0)\psi -\frac{\pi^2 g_0}{\Lambda_0^2}[(\bar\psi\psi)^2+(\bar\psi i\gamma^5 \tau^i \psi)^2].
}
Note that the critical four-Fermi coupling is $g_0{}_\text{c}=1/N_\text{f}N_\text{c}=1/6$.
In Ref.\,\cite{Buballa:2003qv}, the UV cutoff, the four-Fermi coupling constant and the bare quark mass are set to $\Lambda_0=587.9$, $g_0=0.247$ and $m_q=5.6$\,MeV, respectively.
These parameters yield $M=400$\,MeV, $f_\pi=92.4$\,MeV and $m_\pi=135$\,MeV at vanishing temperature and density.
We have reexamined the chiral phase structure with the following initial values: $\Lambda_0=631$\,MeV, $g_0=0.2217$ and $m_q=5.5$\,MeV~\cite{Hatsuda:1994pi}.
In this case, we obtain $M=336$\,MeV, $f_\pi=93$\,MeV and $m_\pi=138$\,MeV at zero temperature and density.
For the present FRG analysis, we used the following initial values as a benchmark point: $\Lambda_0=610\,\text{MeV}$, $g_0=1.3158$.
These values yield the dynamical mass $M=301$\,MeV at $T=\mu=0$.
In the other works with the analytic methods~\cite{Schaefer:2004en,Braun:2011iz,Pawlowski:2014zaa,Schaefer:2007pw,Herbst:2010rf}, the free parameters are set such that one obtains the pion decay constant $f_\pi=93$\,\text{MeV} and the pion mass $m_\pi=138$\,\text{MeV} in the IR regime.\footnote{
The work in \cite{Schaefer:2004en} obtains $f_\pi=87$\,\text{MeV} and $M\sim 300$\,\text{MeV} in the chiral limit.
When the effects of the finite and realistic pion mass are taken into account, the decay constant becomes $f_\pi=93$\,MeV.
}
Refs.\,\cite{deForcrand:2002hgr,Wu:2006su} performed the lattice QCD simulation with imaginary chemical potential $\mu_I=-i\mu$.\footnote{
Ref.\,\cite{deForcrand:2002hgr} uses two-flavor staggered quarks, whereas two-flavor Wilson quarks is used in Ref.\,\cite{Wu:2006su}.
}
By using analytic continuation to real (quark) chemical potential $\mu$, the dependence of the critical temperature on small chemical potential is investigated.
In Ref.\,\cite{Allton:2003vx}, the Taylor-series expansion method is employed.

In Table.\,\ref{tab:comparison}, the (pseudo-)critical temperature $T_\text{c}\fn{\mu=0}$ and the CEP obtained from various models and methods are collected.
We also show the value of the curvature of the chiral phase boundary $\kappa_\mu$ which is defined as 
\al{
\label{curvature of PB}
\frac{T_\text{c}\fn{\mu}}{T_\text{c}\fn{0}}=1-\kappa_\mu \left( \frac{\mu}{\pi T_\text{c}\fn{0}}\right)^2+{\mathcal O}\fn{\left( \frac{\mu}{\pi T_\text{c}\fn{0}}\right)^4}.
}
Note that $\kappa_\mu$ does not depend on the cutoff scale $\Lambda_0$.
We evaluate the curvature in a range $0\leq \mu/(\pi T_\text{c}\fn{0}) \lesssim 0.1$ for the present analysis and the mean-field analysis of the NJL model (III).

\begin{figure}
\begin{center}
\includegraphics[width=100mm]{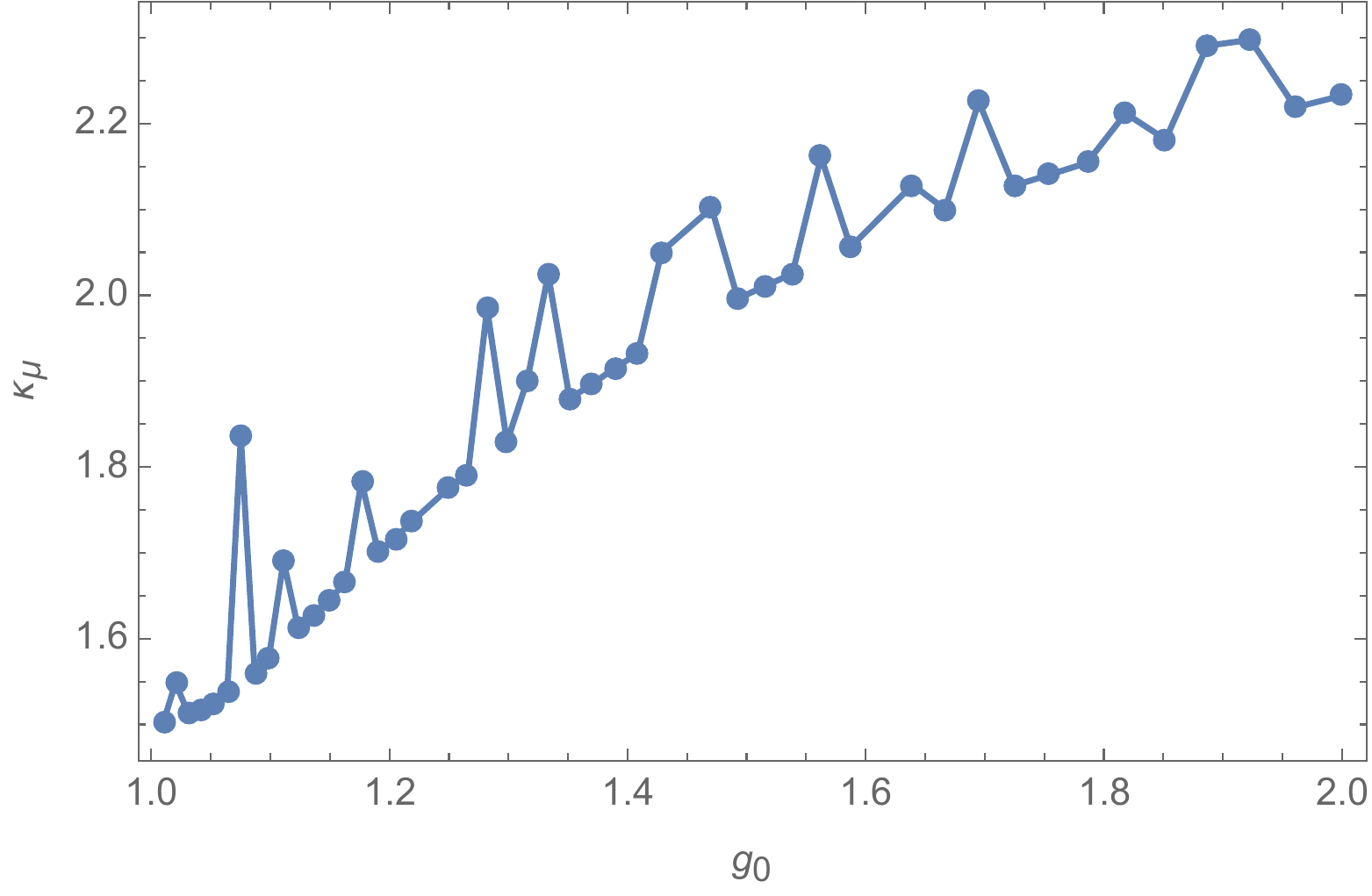}
\end{center}
\caption{The dependence of the curvature $\kappa_\mu$ on the dimensionless four-Fermi coupling $g_0$.}
\label{g0_kappa_mu}
\end{figure}

We see from Table.\,\ref{tab:comparison} that there are no drastic differences in the critical temperature at vanishing chemical potential and the CEP. 
In contrast, the curvatures obtained from our work and the mean-field analysis of the NJL model are somewhat larger than that of the others.
This could be because the mesonic fluctuations are not taken into account in the present FRG and the MFA computations of the NJL model.
Fig.\,\ref{g0_kappa_mu} exhibits the dependence of the curvature $\kappa_\mu$ on the dimensionless four-Fermi coupling $g_0$.
We see that $\kappa_\mu$ tends to be larger with increasing $g_0$.
Note that the value of the critical four-Fermi coupling constant is $g_{0}{}_\text{c}=1$, and as we have seen in the previous subsection, the CEP vanishes below $g_0\simeq 1.2$.
\section{Summary and discussion}\label{summary and discussion}
In this paper, we have studied the chiral phase structure of the NJL model at finite temperature and density using the FRG.
We have seen that when the D$\chi$SB takes place, the solution of the RG equation has to become singular and cannot be evaluated after a certain critical scale.
To overcome this situation, we have introduced the weak solution method to the RG equation.
It has been shown that the weak solution method can appropriately choose the vacuum and describe the chiral phase transition with both second- and first-order.
Previously, in the FRG method, the physical values such as the chiral condensate and the dynamical mass have been evaluated by only introducing the auxiliary field method.
In particular, it was difficult to obtain the chiral phase diagram with the first-order phase transition.
We have shown that the weak solution method allows us to evaluate the chiral dynamics within the pure fermionic system.

Using the FRG with the weak solution method, the D$\chi$SB in QCD at zero temperature and zero density was studied in \cite{Aoki:2012mj}, where effects beyond the ladder approximation (non-ladder effects) on the dynamical mass and the chiral condensate are involved.\footnote{The crossed ladder diagrams and the anomalous dimension of the quark field are included.
These effects cannot be taken into account by the SDE with the ladder approximation and the MFA.
} 
It was shown that the gauge dependence of the physical values is suppressed thanks to including the non-ladder effects.
As a next step, we should investigate the D$\chi$SB in QCD at finite temperature and density.
The weak solution method also can be applied to systems with the color superconductivity and ones in an external magnetic field.\footnote{
See e.g., \cite{Fukushima:2012xw,Mueller:2015fka,Aoki:2015mqa} for the analyses of the system in an external magnetic field using the FRG.
}
In future works, we will apply the FRG with the weak solution method to such systems.

\begin{figure}
\begin{center}
\includegraphics[width=100mm]{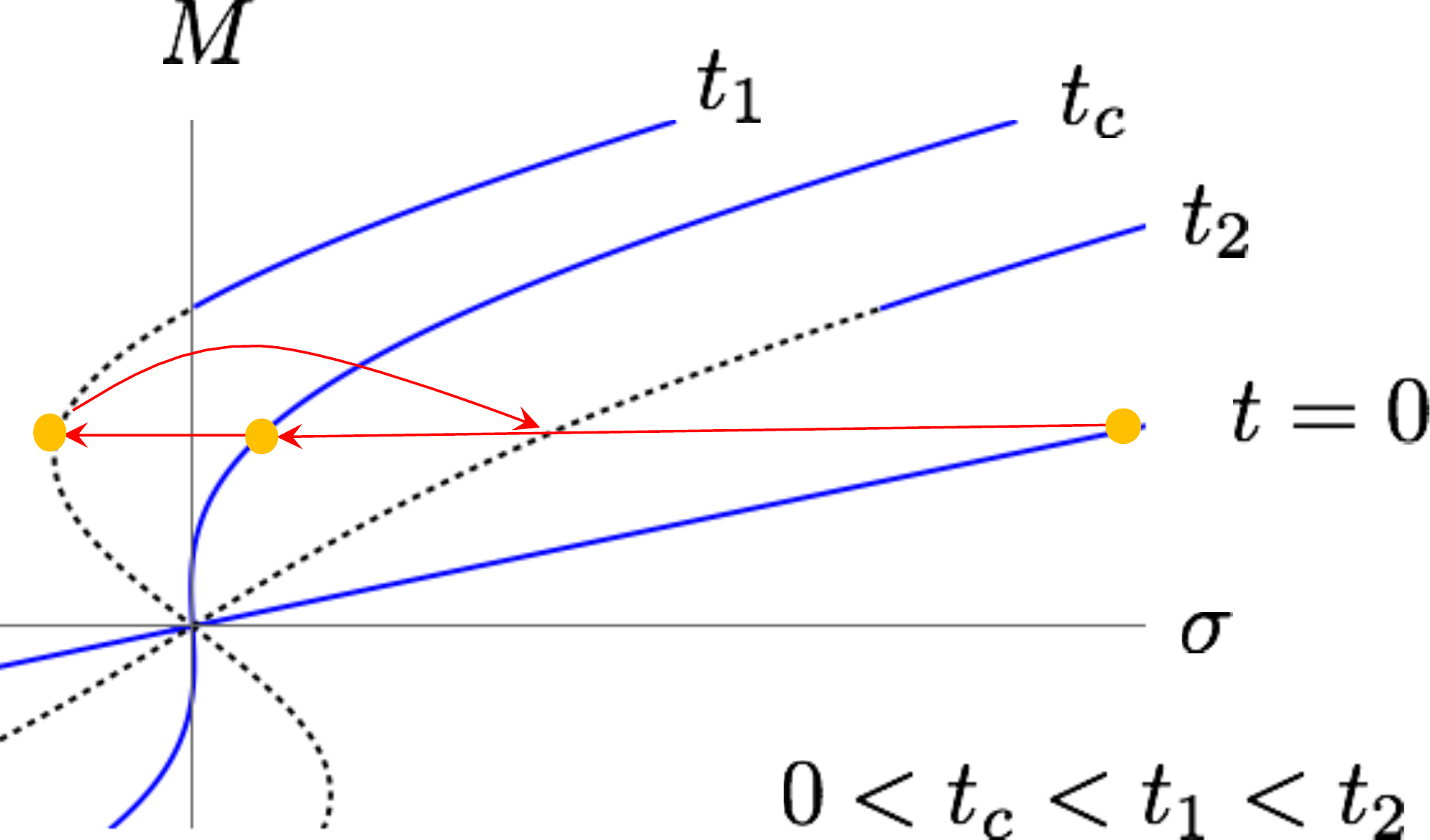}
\end{center}
\caption{Schematic figure for the evolution of the weak solution for the mass function with the optimized cutoff reguralization.
The dotted line is the solution removed by the RH condition.
}
\label{schematic_weak_solution_back}
\end{figure}

Here, we discuss several issues of the weak solution method.
First, we comment on the cutoff scheme and the convexity and concavity of the beta function.
Through this paper, we have analyzed the RG equation obtained by using the WH equation which is formulated with the sharp cutoff regularization.
When the smooth cutoff scheme, e.g., the optimized cutoff regularization~\cite{Litim:2001up}, is employed, we obtain the different form of the beta function for the effective potential; see Eq.~\eqref{optimizedpotential}.
As shown in the appendix~\ref{concavity and convexity}, for the sharp cutoff regularization, the beta function for the fermionic effective potential \eqref{dimensionful RG equation} is concave, i.e., $\p^2 F/\p M^2<0$ for an arbitrary RG scale,  temperature and density.
Therefore, the motion of $\sigma\fn{t}$ described by Eq.~\eqref{cha2} behaves as shown in Fig.~\ref{weak_schematic_motion}.
In contrast, for the case where the optimized cutoff regularization is used, the concavity of the beta function is not guaranteed.
Hence, there are solutions in which the motion of $\sigma\fn{t}$ comes back to the right-hand side as shown in Fig.~\ref{schematic_weak_solution_back}.\footnote{Such a behavior in terms of the four-Fermi coupling constant was reported in \cite{Aoki:2014ova,Aoki:2015hsa}.}
In this case, however, we need some additional information about the solution of the mass function removed once by the RH condition, which corresponds to the dotted line in Fig.~\ref{schematic_weak_solution_back}, in order to define the mass function as a function of $\sigma$.
At present, we do not know how to mathematically define the weak solution for such a ``come-back" solution.\footnote{
It should be stressed here that the ``come-back" solution (physically correct equal area law solution) is a weak solution because it satisfies the RH condition, but it is not a unique solution since there always appear ``rarefaction" solutions in this type of situation. 
We do not even know if we can define some additional condition to assure uniqueness of weak solution in this situation, nor if such condition, if any, may pick up our equal area law solution.
}
Second, let us consider improvements of the approximations.
In this work, we have applied the large-$N$ approximation.
For beyond this approximation, the beta function for the effective potential generally involves its second-order derivative terms with respect to the field $\sigma$, namely $F\fn{M,M',\sigma;t}$ where the prime on the mass function denotes the $\sigma$-derivative and $M=\p_\sigma V$; see Eqs.~\eqref{twopoint1} and \eqref{twopoint2}.
In this case, we cannot naively transfer the $t$- and $\sigma$-derivatives from the mass function to the test function by performing the integration by parts as shown in Eq.~\eqref{weak equation} and then cannot define the weak solution.
The notion of the weak solution that we have introduced in this work corresponds to the so-called ``entropy solution" based on distributions.
On the other hand, it is known that there is also ``viscosity solution"  as a different definition of the weak solution.\footnote{
See e.g., \cite{Evans,Crandallmath} for descriptions on their mathematical definitions.
}
The viscosity solution can be defined for even some second-order nonlinear partial differential equations to which the definition of Eq.~\eqref{weak equation} cannot be applied.
The viscosity solution could be a breakthrough for improving the approximation. 
Therefore, it is important to study applications of the weak solution method in order to analyze critical phenomena in elementary particle physics and condensed matter physics.

\section*{Acknowledgements}
We thank Yasuhiro Fujii and Daisuke Sato for valuable discussions.
The work of K-I. A. is supported by JSPS Grant-in-Aid for Challenging Exploratory Research (No.~16K13848).
The work of M.\,Y. is supported by the DFG Collaborative Research Centre SFB 1225 (ISOQUANT).

\begin{appendix}
\section{Derivation of beta function}\label{derivation of beta function}
In this appendix, we derive the RG equation for the effective potential $V\fn{\bar \psi \psi, \Lambda}$ in both cases of the sharp cutoff and the smooth cutoff schemes.
We derive the RG equation using the WH equation with the sharp cutoff and the Wetterich equation with the smooth cutoff.

\subsection{Sharp cutoff scheme}
For the given effective action~\eqref{effective njl action}, the WH equation reads
\al{\label{njl wegner houghton eqaution}
\frac{\df S_{\rm eff}}{\df t}
=-\frac{1}{2}T\displaystyle \sum_{n=-\infty}^\infty \int \frac{\df ^{3}p}{(2\pi)^{3}}\,\Lambda\delta\fn{|\vec p|-\Lambda}\,
{\rm tr}\log\fn{ \frac{\overrightarrow \delta}{\delta \phi_p^{\rm T}} S_{\rm eff}\frac{\overleftarrow \delta}{\delta \phi_{-p} }},
}
where $\phi^{\rm T} _p=\left(\psi^{\rm T}\fn{-p}, \bar \psi \fn{p}\right)$.
To derive the RG equation, we introduce the mean-field defined as
\al{
\psi\fn{p}&= (2\pi)^4 \delta^{(4)}\fn{p}\, \psi,&
\bar \psi \fn{p}&= (2\pi)^4 \delta^{(4)}\fn{p}\, \bar \psi.&
}
We apply them after evaluating the functional derivatives of $S_{\rm eff}$ with respect to $\phi\fn{p}$.
The inverse two point function is given by
\al{\label{full inverse two point function}
\frac{\overrightarrow \delta}{\delta \phi_p^{\rm T}} S_{\rm eff}\frac{\overleftarrow \delta}{\delta \phi_{p'} }
=\pmat{
-\displaystyle \frac{\overrightarrow \delta}{\delta \psi^{\rm T}}V\frac{\overleftarrow \delta}{\delta \psi }
& -i(\Slash p -i\mu \gamma_0)^{\rm T} - \displaystyle \frac{\overrightarrow \delta}{\delta \psi^{\rm T}}V\frac{\overleftarrow \delta}{\delta \bar \psi ^{\rm T}}\\[15pt]
 -i(\Slash p +i\mu\gamma_0)- \displaystyle \frac{\overrightarrow \delta}{\delta \bar\psi}V\frac{\overleftarrow \delta}{\delta \psi } 
&
-\displaystyle \frac{\overrightarrow \delta}{\delta \bar \psi}V\frac{\overleftarrow \delta}{\delta \bar\psi^{\rm T} }
}(2\pi)^4\delta^{(4)} \fn{ p-p'}
,
}
where we can calculate
\al{
\frac{\overrightarrow \delta}{\delta \psi_i^{\rm T}}V\frac{\overleftarrow \delta}{\delta \psi_j}
&= -{\bar \psi}_i^\text{T}{\bar \psi}_j (\p_\sigma^2 V),&
\displaystyle \frac{\overrightarrow \delta}{\delta \bar \psi_i}V\frac{\overleftarrow \delta}{\delta \bar\psi_j^{\rm T}} 
&= -\psi_i \psi_j^\text{T} (\p_\sigma^2 V),&
\label{twopoint1}
\\
\displaystyle \frac{\overrightarrow \delta}{\delta \bar\psi_i}V\frac{\overleftarrow \delta}{\delta \psi_j} 
&= (\p_\sigma V)\delta_{ij} +\psi_i{\bar \psi}_j (\p_\sigma^2 V),&
\displaystyle \frac{\overrightarrow \delta}{\delta \psi_i^{\rm T}}V\frac{\overleftarrow \delta}{\delta \bar \psi_j^{\rm T}}
&= -(\p_\sigma V)\delta_{ij} - (\psi_j {\bar \psi}_i)^\text{T} (\p_\sigma^2 V).
\label{twopoint2}
}
Here we employ the following approximation:
\al{
\displaystyle \frac{\overrightarrow \delta}{\delta \psi^{\rm T}}V\frac{\overleftarrow \delta}{\delta \psi } &= 0,&
\displaystyle \frac{\overrightarrow \delta}{\delta \bar \psi}V\frac{\overleftarrow \delta}{\delta \bar\psi^{\rm T}} &= 0,&
\displaystyle \frac{\overrightarrow \delta}{\delta \bar\psi}V\frac{\overleftarrow \delta}{\delta \psi } 
&= \p_\sigma V,&
\displaystyle \frac{\overrightarrow \delta}{\delta \psi^{\rm T}}V\frac{\overleftarrow \delta}{\delta \bar \psi ^{\rm T}}
&= -\p_\sigma V,&
}
which corresponds to the so-called large-$N$ leading approximation ($N_\text{c}\to\infty$).

The matrix~\eqref{full inverse two point function} is reduced to the simple form,
\al{\label{simple inverse two point function}
\frac{\overrightarrow \delta}{\delta \phi_p^{\rm T}} S_{\rm eff}\frac{\overleftarrow \delta}{\delta \phi_{p'} }
\simeq
\pmat{
0 & -i\Slash p^-{}^{\rm T} + \p_\sigma V \\[15pt]
 -i\Slash p^+ - \p_\sigma V
&
0
}(2\pi)^4\delta^{(4)} \fn{ p-p'},
}
where we defined $\Slash p^\pm:=\Slash p \pm i\mu\gamma_0$.
After using the formula of matrix, ${\rm tr}\,\log\fn{A} =\log{\rm det}\,\fn{A}$, and ${\rm tr}\log\fn{\Slash p^\pm \mp i\p_\sigma V}=2\log\fn{\left(p^\pm\right)^2+(\p_\sigma V)^2}$, the shell momentum integral for Eq.~\eqref{njl wegner houghton eqaution} is evaluated as follows:
\al{
&\int \frac{\df ^3p}{(2\pi)^3}\, \Lambda\delta\fn{|\vec p|-\Lambda}{\rm tr}\log\fn{ \frac{\overrightarrow \delta}{\delta \phi_p^{\rm T}} S_{\rm eff}\frac{\overleftarrow \delta}{\delta \phi_{-p} }}\nn
&= \frac{4\pi\Omega\Lambda}{(2\pi)^3}
 \int \df |\vec p|\,\delta\fn{|\vec p|-\Lambda}|\vec p|^2 \,\nn
&\qquad \times \bigg\{ 2\log\left[ (p_0 + i\mu)^2 + |\vec p|^2 +(\p_\sigma V)^2 \right]
+2\log\left[(p_0 - i\mu)^2 +|\vec p|^2 + (\p_\sigma V)^2 \right]
\bigg\}
\nn
&=\Omega\frac{\Lambda^3}{\pi^2}
\left[
\log\left[ p_0^2 + (E+\mu)^2\right] + \log\left[p_0^2 + (E-\mu)^2\right]
\right]\nn
&=\Omega\frac{\Lambda^3}{\pi^2}
\bigg[
\int ^{(E+\mu)^2}_{T^2}\frac{\df \theta^2}{\theta^2 +p_0^2}
+\int ^{(E-\mu)^2}_{T^2}\frac{\df \theta^2}{\theta^2 +p_0^2}
+C_n
\bigg],
\label{evaluation1}
}
where $E:=\sqrt{\Lambda^2 +(\p_\sigma V)^2}$ and $\Omega:= (2\pi)^4\delta^{(4)}\fn{0}$ is a volume of space-time.
Here $C_n:= 2\log \fn{ T^2 +p_0^2}$ does not depend on the effective potential and hereafter it is ignored.

Next, we evaluate the Matsubara summation using the formula,
\al{
\sum_{n=-\infty}^\infty \frac{1}{(2n+1)^2\pi^2 +\beta^2\theta^2}
&=\frac{1}{\beta\theta}\left( \frac{1}{2} -\frac{1}{\e^{\beta\theta}+1} \right)\nn
&=\frac{1}{2\beta\theta}\tanh \fn{\frac{\beta \theta}{2}}
.\label{formula of summation}
}
The summation for Eq.~\eqref{evaluation1} with $p_0=(2n+1)\pi T$ becomes
\al{
&\Omega\frac{\Lambda^3}{\pi^2}T\displaystyle \sum_{n=-\infty}^\infty\bigg[
\int ^{(E+\mu)^2}_{T^2}\frac{\df \theta^2}{\theta^2 +p_0^2}
+\int ^{(E-\mu)^2}_{T^2}\frac{\df \theta^2}{\theta^2 +p_0^2}
+C_n
\bigg]\nn
&\qquad =\Omega\frac{\Lambda^3}{\pi^2}
\bigg[
\int ^{E+\mu}_{T}\left( 1 -\frac{2}{\e ^{\beta\theta}+1}\right)\df \theta
+\int ^{E-\mu}_{T}\left( 1 -\frac{2}{\e ^{\beta\theta}+1}\right)\df \theta
\bigg]\nn
& \qquad =\Omega\frac{2\Lambda^3}{\pi^2}
\bigg[ 
E + T\log \fn{1+\e^{-\beta(E-\mu)}}
+T\log \fn{1+\e^{-\beta(E+\mu)}}
+ C
\bigg],
}
where $C:=-2T +2T\log \fn{1 + \e^{-1}}$ is a constant value and can be ignored.
To summarize, using $\p_tV =-\p_tS_{\rm eff}/\Omega$, we obtain the RG equation for the effective potential as follows:
\al{\label{deriven dimensionful RG equation}
\p_t V\fn{\sigma;t}= \frac{\Lambda^3}{\pi^2}
\bigg[
E +T\log \fn{1+\e^{-\beta (E-\mu)}} +T \log \fn{1+\e^{-\beta(E+\mu)}}
\bigg].
}

\subsection{Smooth cutoff scheme}
To obtain the RG equation in case of the smooth cutoff scheme, we use the Wetterich equation~\cite{Wetterich:1992yh,Morris:1993qb},
\al{\label{full wetterich equation}
\frac{\df \Gamma_{\Lambda}}{\df t}=\frac{1}{2}\int \frac{\df ^4p}{(2\pi)^4}\,
{\rm str}\left[ \frac{\overrightarrow\delta }{\delta \phi_p}\Gamma_{\Lambda}\frac{\overleftarrow\delta }{\delta \phi_{-p}}+{\mathcal R}_{\Lambda} \right]^{-1}\cdot \left( \p_t {\mathcal R}_\Lambda\right),
}
where the $\Gamma_{\Lambda}$ is the Legendre effective action and $\mathcal R_\Lambda$ is the cutoff profile function.
See e.g. \cite{Gies:2006wv,Oda:2015sma} for its explicit derivation.

For the level of approximation at present, the Legendre effective action is equivalent to the Wilsonian effective action~\cite{Morris:1993qb}, thus it is given by
\al{\label{legendre effective action}
\Gamma_\Lambda[\psi, \bar\psi]
=
\int \df^4 x\left[\bar\psi \Slash \p \psi
 -V\fn{\bar\psi \psi;\Lambda}\right].
}
Therefore the two point function $\frac{\overrightarrow\delta }{\delta \phi_p}\Gamma_{\Lambda}\frac{\overleftarrow\delta }{\delta \phi_{-p}}$ becomes the same form as Eq.~\eqref{simple inverse two point function}.
As mentioned above Eq.~\eqref{shell mode 3d}, since we insert the cutoff scale $\Lambda$ into the momentum in the spatial direction, the cutoff profile function is given by
\al{
{\mathcal R}_\Lambda\fn{\vec p,\vec p'}=
\pmat{
0 &i\vec {\Slash p}{}^{\rm T}\, r_\Lambda\fn{\vec p^2/\Lambda^2}\\
i\vec{\Slash p}\, r_\Lambda\fn{\vec p^2/\Lambda^2} & 0
}(2\pi)^4 \delta^{(4)}\fn{p-p'}.
}
We see
\al{
\frac{\overrightarrow\delta }{\delta \phi_p}\Gamma_{\Lambda}\frac{\overleftarrow\delta }{\delta \phi_{p'}}
+\mathcal R_\Lambda\fn{p,p'}
=\pmat{
0 & -i \Slash p^- +\p_\sigma V\\
-i \Slash p^+ -\p_\sigma V & 0
}(2\pi)^4 \delta^{(4)}\fn{p-p'},
}
where we defined $p^\pm:=\left( p_0\pm i\mu,\, \vec {\Slash p}(1+r_\Lambda) \right)$, and used the large-$N$ approximation.
Here we employ the optimized cutoff function~\cite{Litim:2001up,Litim:2006ag} as the smooth cutoff scheme, namely,
\al{
r_{\Lambda}\fn{|{\vec p}|^2/\Lambda^2}=\left(\sqrt{ \frac{\Lambda^2}{|\vec p|^2}}-1 \right)\, \theta\fn{1-\frac{|\vec p|^2}{\Lambda^2}}.
}

Let us now compute the beta function for the effective potential.
For the system~\eqref{legendre effective action} at finite temperature and density, the Wetterich equation~\eqref{full wetterich equation} reads
\al{\label{wetterich equation}
\frac{\df \Gamma_{\Lambda}}{\df t}
=-\frac{1}{2}T\displaystyle \sum_{n=-\infty}^\infty\int \frac{\df ^3p}{(2\pi)^3}\,
\tilde \p_t\,{\rm tr}\log\left[ \frac{\overrightarrow\delta }{\delta \phi_p}\Gamma_{\Lambda}\frac{\overleftarrow\delta }{\delta \phi_{-p}}+{\mathcal R}_{\Lambda} \right],
}
where $\phi^{\rm T} \fn{p}=\left(\psi^{\rm T}\fn{-p}, \bar \psi \fn{p}\right)$, and we introduced the derivative with respect to $t$, which acts on only the cutoff dependence of $\mathcal R_\Lambda$, thus, it is $\tilde \p_t:=\frac{\df \mathcal R_\Lambda}{\df t}\cdot \frac{\p}{\p  \mathcal R_\Lambda}=\frac{\p r_\Lambda}{\p t}\cdot \frac{\p}{\p  r_\Lambda}$.
To compare this form with the WH equation~\eqref{njl wegner houghton eqaution},
we notice that both equations are almost the same structure and that $\tilde \p_t$ plays the role of taking the shell momentum mode.
Thereby, we can use the same calculations as the case of the sharp cutoff scheme, and thus we evaluate the RHS of Eq.~\eqref{wetterich equation} as follows:
\al{
({\rm RHS})&=-\frac{\Omega}{2}T\displaystyle \sum_{n=-\infty}^\infty\int \frac{\df ^3p}{(2\pi)^3}\,
\tilde \p_t\, \bigg\{ 
2\log\left[ (p_0+i\mu)^2 +  |\vec p|^2(1+r_\Lambda)^2+(\p_\sigma V)^2\right] 
\nn
&\phantom{-\frac{1}{2}T\displaystyle \sum_{n=-\infty}^\infty\int \frac{\df ^3p}{(2\pi)^3}\,
\tilde \p_t\, \frac{1}{2}\bigg\{ }
\quad+ 2\log\left[(p_0-i\mu)^2 +|\vec p|^2(1+r_\Lambda)^2+(\p_\sigma V)^2\right]
\bigg\}\nn
& =-\Omega T\displaystyle \sum_{n=-\infty}^\infty\int \frac{\df ^3p}{(2\pi)^3}\frac{\p r_\Lambda}{\p t}
\bigg\{ 
\frac{2|\vec p|^2(1+r_\Lambda)}{ (p_0+i\mu)^2 +  |\vec p|^2(1+r_\Lambda)^2+(\p_\sigma V)^2}\nn
&\phantom{-T\displaystyle \sum_{n=-\infty}^\infty\int \frac{\df ^3p}{(2\pi)^3}\frac{\p r_\Lambda}{\p t}
\bigg\{ }
\quad +\frac{2|\vec p|^2(1+r_\Lambda)}{ (p_0-i\mu)^2 +  |\vec p|^2(1+r_\Lambda)^2+(\p_\sigma V)^2}
\bigg\}\nn
&=-\frac{\Omega}{2\pi^2}T\displaystyle \sum_{n=-\infty}^\infty\int 
\df|\vec p|\, |\vec p|^2 \left( -\frac{\Lambda}{|\vec p|} \theta\fn{1-|\vec p|^2/\Lambda^2}\right)(2|\vec p|^2(1+r_\Lambda))\nn
&\quad \times
\bigg\{
\frac{1}{ (p_0+i\mu)^2 +  |\vec p|^2(1+r_\Lambda)^2+(\p_\sigma V)^2}
+\frac{1}{ (p_0-i\mu)^2 +  |\vec p|^2(1+r_\Lambda)^2+(\p_\sigma V)^2}
\bigg\}\nn
&=\frac{\Omega\Lambda^5}{3\pi^2}T\displaystyle \sum_{n=-\infty}^\infty
\bigg\{
\frac{1}{ (p_0+i\mu)^2  + E^2}
+\frac{1}{ (p_0-i\mu)^2 +  E^2}
\bigg\}\nn
&=\frac{\Omega\Lambda^5}{3\pi^2}T\displaystyle \sum_{n=-\infty}^\infty \frac{1}{E}
\bigg\{
\frac{E+\mu}{ p_0^2 + (E+\mu)^2}
+\frac{E-\mu}{ p_0^2 + (E-\mu)^2}
\bigg\}\nn
&=
\frac{\Omega\Lambda^5}{6\pi^2E}\,
\left[ \tanh \fn{ \frac{E+\mu}{2T} } 
+ \tanh \fn{ \frac{E-\mu}{2T} }
\right],
}
where we used the second line of Eq.~\eqref{formula of summation} for the Matsubara summation.
Then we obtain the RG equation,
\al{
\p_t V\fn{\sigma; t}=-\frac{\Lambda^5}{6\pi^2E}
\left[
\tanh \fn{ \frac{E+\mu}{2T}} + \tanh \fn{ \frac{E-\mu}{2T}}
\right].
\label{optimizedpotential}
}

\section{Concavity and convexity of beta function}\label{concavity and convexity}
We see that the beta function for the fermionic effective potential is concavity.
To this end, we first consider the second-order derivative of the beta function for the sharp cutoff regularization case (Eq.~\eqref{deriven dimensionful RG equation}) with respect to $M$.
Here we define $N_-:=\e^{\beta (E-\mu)}$ and $N_+:=\e^{\beta(E+\mu)}$.
Then the beta function is 
\al{
F\fn{M;t}:= -\frac{\Lambda^3}{\pi^2}
\bigg[
E +T\log \fn{1+N_-^{-1}} +T \log \fn{1+N_+^{-1}}
\bigg].
}
We have
\al{
f\fn{M;t}&:=
\frac{\p^2}{\p M^2}\left[E+T(\log \fn{1+N_-^{-1}}+\log \fn{1+N_+^{-1}})\right]
\nn
\quad
&=E''
+\left( \frac{N_-E'}{(1+N_-)^2} - \frac{E''}{1+N_-} \right)
+\left( \frac{N_+E'}{(1+N_+)^2} - \frac{E''}{1+N_+} \right)\nn
&=\frac{N_- E'}{(1+N_-)^2} + \frac{N_+E'}{(1+N_+)^2}
+\frac{E''(N_+N_--1)}{(1+N_+)(1+N_-)}.
}
Since 
\al{
E':=\frac{\p E}{\p M}&= \frac{M}{E}>0,&
E'':=\frac{\p^2 E}{\p M^2}&= \frac{\Lambda^2}{E^{3/2}}>0,&
N_-&>0, &
N_+&>0, &
N_+N_- -1&>0,&
}
we find $f\fn{M;t}>0$ and then $\p^2F\fn{M;t}/\p M^2=-\Lambda^3 f\fn{M;t}/\pi^2<0$.
Hence, the beta function at a mass function for arbitrary temperature and density is concave as a function of $M$.

\begin{figure}
\begin{center}
\includegraphics[width=140mm]{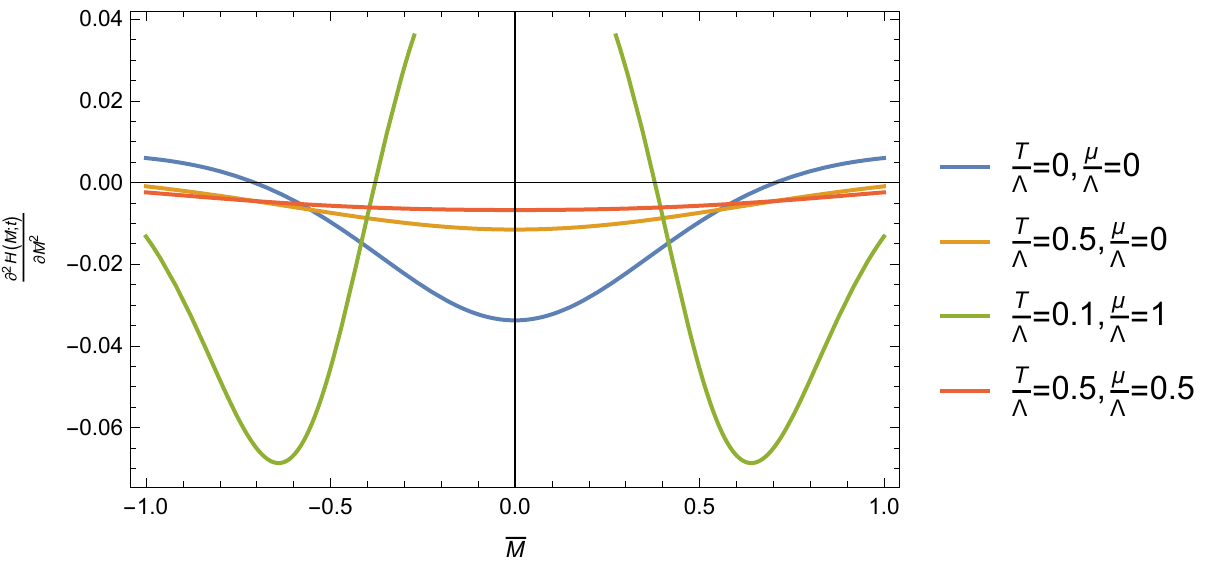}
\end{center}
\caption{The dimensionless $\p^2 \bar H\fn{{\bar M};t}/\p {\bar M}^2$ as a function of $\bar M$.}
\label{H_func}
\end{figure}

Next, the case for the optimized cutoff regularization is considered.
The beta function is
\al{
H\fn{M;t}:=
\frac{\Lambda^5}{6\pi^2E}
\left[
\tanh \fn{ \frac{E+\mu}{2T}} + \tanh \fn{ \frac{E-\mu}{2T}}
\right].
}
Its second-order derivative with respect to $M$ is
\al{
\frac{\p^2 H}{\p M^2}&=\frac{\Lambda^5}{6\pi^2}\Bigg[
-\frac{1}{TE}\left(\frac{E'^2}{E} -\frac{E''}{2}\right)\left( \text{sech}^2\fn{\frac{E-\mu}{2T}}+\text{sech}^2\fn{\frac{E+\mu}{2T}} \right)\nn
&\quad
-\frac{E'^2}{2T^2E}\left( \text{sech}^2\fn{\frac{E-\mu}{2T}}\tanh\fn{\frac{E-\mu}{2T}}
 +\text{sech}^2\fn{\frac{E+\mu}{2T}}\tanh\fn{\frac{E+\mu}{2T}} \right)\nn
&\quad
+\frac{2}{E^2}\left(\frac{E'^2}{E}-\frac{E''}{2} \right)\left(\tanh \fn{ \frac{E+\mu}{2T}} + \tanh \fn{ \frac{E-\mu}{2T}} \right)
\Bigg].
\label{delHM}
}
Since one cannot analytically see its convexity and concavity, we plot it as a function of $M$ in Fig.~\ref{H_func}, where the dimensionless $\p^2 \bar H\fn{{\bar M};t}/\p {\bar M}^2=(\p^2 H\fn{{M}/\Lambda;t}/\p M^2)/\Lambda^2$ is plotted.
We see that the concavity of the beta function in the optimized cutoff regularization case is not guaranteed since Eq.~\eqref{delHM} can have positive values.
\end{appendix}

\bibliography{refs}

\end{document}